\documentclass[journal]{IEEEtran}
\usepackage{amsmath,amsfonts}
\usepackage{algorithmic}
\usepackage{algorithm}
\usepackage{array}
\usepackage[caption=false,font=normalsize,labelfont=sf,textfont=sf]{subfig}
\usepackage{textcomp}
\usepackage{stfloats}
\usepackage{url}
\usepackage{verbatim}
\usepackage{graphicx}
\usepackage{cite}
\usepackage{booktabs}
\usepackage{longtable}
\usepackage{amsmath,amssymb,amsfonts}
\usepackage{amsthm}
\usepackage{makecell}
\usepackage{upgreek}
\theoremstyle{definition}
\newtheorem{definition}{Definition}
\usepackage{setspace}

\begin{document}

\title{Learning Wireless Data Knowledge Graph for Green Intelligent Communications: Methodology and Experiments}

\author{Yongming~Huang,~\IEEEmembership{Senior Member,~IEEE},
      Xiaohu~You,~\IEEEmembership{Fellow,~IEEE},
      Hang~Zhan,
      Shiwen~He,~\IEEEmembership{Member,~IEEE},
      Ningning~Fu,~\IEEEmembership{Graduate Student Member, IEEE},
      and~Wei Xu,~\IEEEmembership{Senior Member,~IEEE}}







\markboth{Journal of \LaTeX\ Class Files,~Vol.~14, No.~8, August~2024}%
{Huang \MakeLowercase{\textit{et al.}}: Towards Green and Lightweight Intelligent Communications: Pervasive Multi-level Native AI Architecture and Methodology Based on Wireless Data Knowledge Graph}


\maketitle
\begin{abstract}
Intelligent communications have played a pivotal role in shaping the evolution of 6G networks. Native artificial intelligence (AI) within green communication systems must meet stringent real-time requirements. To achieve this, deploying lightweight and resource-efficient AI models is necessary. However, as wireless networks generate a multitude of data fields and indicators during operation, only a fraction of them imposes significant impact on the network AI models. Therefore, real-time intelligence of communication systems heavily relies on a small but critical set of the data that profoundly influences the performance of network AI models. However, this aspect remains unclear and often overlooked. These challenges underscore the need for innovative architectures and solutions. In this paper, we propose a solution, termed the pervasive multi-level (PML) native AI architecture, which integrates the concept of knowledge graph (KG) into the intelligent operational manipulations of mobile networks, resulting in the establishment of a wireless data KG. Leveraging the wireless data KG, we characterize the massive and complex data collected from wireless communication networks and analyze the relationships among various data fields. The obtained graph of data field relations enables the on-demand generation of minimal and effective datasets, referred to as feature datasets, tailored to specific application requirements. Additionally, this approach facilitates the removal of redundant data fields with minimal impact on network AI performance. Consequently, this architecture not only enhances AI training, inference, and validation processes but also significantly reduces resource wastage and overhead for communication networks. To implement this architecture, we have developed a specific solution comprising a \textbf{s}patio-\textbf{t}emporal hete\textbf{r}ogeneous graph att\textbf{e}ntion neur\textbf{a}l network \textbf{m}odel (STREAM) as well as a feature dataset generation algorithm. Experiments are conducted to validate the effectiveness of the proposed architecture. The first experiment validates the advantages of STREAM in the wireless data KG link prediction, demonstrating its exceptional capability in handling the spatio-temporal data. The second experiment confirms that the PML native AI architecture effectively reduces data scale and computational costs of AI training by almost an order of magnitude. This affirms its potential to support green and prompt-response network intelligence for the next-generation wireless networks.
\end{abstract}


\section{Introduction}

\IEEEPARstart{T}{he} future landscape of mobile networks is undergoing rapid expansion, characterized by a surge growth in connected devices, mobile data traffic, and an imperative for new functionalities and applications \cite{ITU2023}. Consequently, forthcoming networks are expected to embrace innovative architectures and supporting technologies to ensure the extreme connectivity for seamless coverage and high-value services \cite{You2021Wang}. Traditional operational models and rule-based algorithms confront challenges in adapting to evolving user demands and network environments. Though it is widely known that achieving native AI is crucial to enable advanced autonomous driving and customized services within the network \cite{Masood2023Farooq}, the development of native AI  driven by data and model synergy in wireless networks is still in its early stages, facing significant challenges in data, architecture, and algorithm design \cite{Letaief2019Chen}. One specific challenge lies in real-time requirements for native AI in communication systems \cite{Chen2020Liu}. Leveraging rapidly advancing large language models (LLMs) can be helpful at the cost of extensive computational and storage resources, hindering real-time communication and exacerbating energy consumption. According to the GSMA report, considering only mobile networks, the annual energy consumption is approximately 130 TWh, with greenhouse gas emissions of around 110 MtCO2e, accounting for about 0.6\% of global electricity consumption and 0.2\% of global greenhouse gas emissions. As per the International Energy Agency's ``Net Zero by 2050" report, global greenhouse gas emissions need to be cut in half by 2030 \cite{IEA}. Therefore, the ``green" issue will continue to be a key focus in the development of 6G \cite{Huang2019Yang}. In future 6G intelligent communication, the development of green and lightweight intelligent solutions will be especially critical.

Among these challenges, data stands out as the cornerstone forming the crucial foundation \cite{Polese2021Jana}. One primary way of attaining green and lightweight native AI primarily lies in understanding the data comprehensively, extracting highly-valuable knowledge, and unveiling essential data insights through a meticulous process of data analysis and exploration. Mobile communication networks generate tons of data fields and indicators during their network operations. Among the massive amount of data, certain data fields and indicators have interdependent effects on AI models, while others poses minimal impact. Hence, the effective classification, analysis, and extraction of features from diverse data types, along with geneating minimal and effective datasets (referred to as feature datasets) tailored for different on-demand applications, is crucial for driving AI training, inference, and validation. This process stands out as the most fundamental challenge in the development of 6G native AI and represents the most efficient approach to achieving intelligent and simplified networks \cite{You2023Huang}.

To address these challenges, we advocate a new architecture of pervasive multi-level (PML) native AI for networks by involving the proposed  knowledge graphs (KG) into the domain of mobile networks, resulting in the establishment of a wireless data KG. The core of this architecture lies in the utilization of the wireless data KG to organize and  condense intricate and disordered wireless data, thereby extracting a concise subset of the wireless data that represents the most effective and critical impact on network AI models using a large volume of wireless data. As a result, this approach loosens the need for extensive dataset scale that is traditionally required for the AI model training, consequently reducing the costs associated with training these models. This ultimately leads to the creation of a green, efficient, and lightweight AI network.

\subsection{Related Work}


In recent years, there has been a surge in the development of native AI architectures tailored for wireless networks, which has enhanced the performance of wireless systems in both academia and industry. Researchers have developed data-driven architectures and methodologies for managing wireless data, incorporating deep learning (DL) techniques and intelligent computing frameworks \cite{Xu2023, Xu2018Xu, Liu2024AAAI}. Additionally, other researchers have explored the general processes involved in handling wireless big data, encompassing data acquisition, preprocessing, storage, model design, training, and application \cite{Liu2020Bi}. It is important to note that the aforementioned studies predominantly focus on leveraging data and AI algorithms to address existing challenges within wireless networks, without delving deeply into comprehensive analysis and understanding of the system itself. Moreover, while these endeavors have introduced new data processing technologies into the domain of wireless communication, the potential requirement of additional overhead and energy consumption stemming from these technologies has not been adequately considered. Therefore, the proposed PML native AI architecture not only enables the utilization of the wireless data KG to elucidate the underlying relationships within wireless data but also facilitates the generation of feature datasets through intelligent inference. This approach effectively reduces the data collection scale and the training cost of AI models.

The core component of the PML native AI architecture is to construct a high-quality wireless data KG. Currently, wireless data KGs are typically crafted  by experts through parsing the parsing of the 3GPP protocols. However, this manual construction process is labor-intensive and prone to information loss and even errors due to the subjectivity and limitations of expert knowledge. Moreover, the unpredictable, intricate, and dynamic nature of future networks transforms the wireless data KG into a massive and highly dynamic KG for each communication instance. Hence, achieving a balance between the quality, efficiency, and cost in constructing wireless data KGs has become a fundamental concern in practice. To enhance the efficiency and accuracy of establishing a wireless data KG, it is imperative to integrate wireless expert knowledge and protocol understanding with the wireless big data, fully exploring and utilizing their potential. Consequently, task of the link prediction based on wireless big data and wireless data KGs emerges as a key research focus. Traditional link prediction algorithms only utilize graph structures and attribute information to calculate the similarity between nodes \cite{Ding2022Lai}. In the wireless data KGs, however, nodes not only possess graph structure and attribute data but are also accompanied by collected wireless big data. Moreover, relationships within the wireless data KG, as well as the data from nodes, exhibit variability under different environmental conditions. This results in specific instantiations of the wireless data KG at each sampling point and contributes to a highly dynamic framework. Nodes within each instantiation of the wireless data KG not only reveal spatial correlations related to protocol specification processes but also demonstrate temporal correlations across different wireless data KG instantiations. Given these characteristics, conventional link prediction methods are not directly applicable to the wireless data KG. Consequently, exploring the comprehensive integration of wireless big data, graph attributes, and graph structure data becomes essential. The development of appropriate graph embedding algorithms and their applications in the link prediction for wireless data KG management is thus imperative.

Graph embedding is a method that reshapes complex graph data into a continuous low-dimensional space. This process preserves vital information, capturing the inherent network structure while efficiently compressing redundant data \cite{Shen2022Zhang}. Across diverse domains, e.g., bioinformatics and social networks, graph embedding methods have been successful in seeking to reveal hidden relationships and features within graph data. Despite their adaptability, these methods frequently overlook the nuanced manipulations of the node-level data, neglecting the dynamic relationships inherent in the graph. In addition, they often failed to account for the non-uniform nature of node attributes and the robust spatio-temporal correlations within the collected data \cite{Wang2017Mao}.

Upon completing the graph embedding learning and link prediction tasks, we not only acquire the graph structure of the wireless data KG but also determine the similarity between nodes, which provides a metric for the relationship between nodes. To uncover the  critical factors that influence the Key Performance Indicators (KPIs) in experiments, we exploit the graph structure and the degree of inter-node relationships to evaluate the impact of each node on the KPIs and rank them accordingly. Subsequently, by considering both fitness and feature compression rate, we can choose the minimal efficient dataset comprising the top-ranked nodes that are identified with the most significant impact on KPIs. This procedure ensures lightweight input for subsequent AI algorithms, enabling real-time and green intelligence.

\subsection{Contributions}

Based on the aforementioned considerations, this paper proposes a PML native AI architecture that utilizes a wireless data KG as its core component, contributing to the advancement of green intelligence. By extracting a minimal yet highly effective feature dataset closely connected to the network AI performance from massive wireless data, this architecture supports subsequent lightweight AI models, thereby reducing computational costs. Firstly, a wireless data KG embedding learning model referred to as the \textbf{S}patio-\textbf{T}emporal Hete\textbf{r}ogeneous Graph Att\textbf{e}ntion Neur\textbf{a}l Network \textbf{M}odel (STREAM) is introduced. Secondly, precise degrees of association between wireless data fields and the graph structure obtained through STREAM is utilized to generate the feature dataset. Finally, the effectiveness of the generated feature dataset is validated through a experiment. Technical contributions of this work are summarized below.

\begin{itemize}
    \item We establish a PML native AI architecture that leverages a wireless data KG as its core component, extracting crucial and effective feature datasets from massive and complex wireless big data. This approach significantly diminishes the data volume needed by conventional AI model training, thereby promoting a green, real-time, and lightweight AI solution for the wireless network.

   \item We develop a novel end-to-end STREAM framework specifically tailored to the discovered characteristics of wireless data KG. This framework excels in extracting heterogeneous spatial, temporal, and attribute information from wireless networks across various operating states. The STREAM is verified skilled in link prediction tasks, enabling a more precise capture of the correlations underlying wireless data fields in dynamically complicated communication environments.  It consequently promotes more accurate and intelligent construction and refinement of the wireless data KG. These characteristics have been validated through extensive experiments, which demonstrate superior performance compared to existing alternative methods.

   \item We propose a method for generating feature datasets based on the wireless data KG and the two evaluation metrics for assessing feature datasets. The proposed method offers a benchmark for identifying the minimal yet effective dataset with the dominating impact on the performance of network AI. Experimental validations have also demonstrated that the obtained feature dataset can significantly reduce the costs, thereby providing a practical pathway for realizing green intelligent  wireless networks.

\end{itemize}

The remainder of this paper follows the following structure: Section II introduces a PML  native AI framework based on the wireless data KG, detailing the definition and characteristics of the wireless data KG, along with an illustrative example. Section III provides a detailed exposition of the construction and application of the wireless data KG. Section IV presents specific techniques for constructing the wireless data KG with a blend of knowledge and data, as well as methods for generating feature datasets using the wireless data KG. Section V encompasses the experimental setup and results. Finally, Section VI concludes the paper and discusses future research directions.

\section{Wireless data KG based PML native AI architecture}

\begin{figure}[!t]
	\includegraphics[width=3.7in]{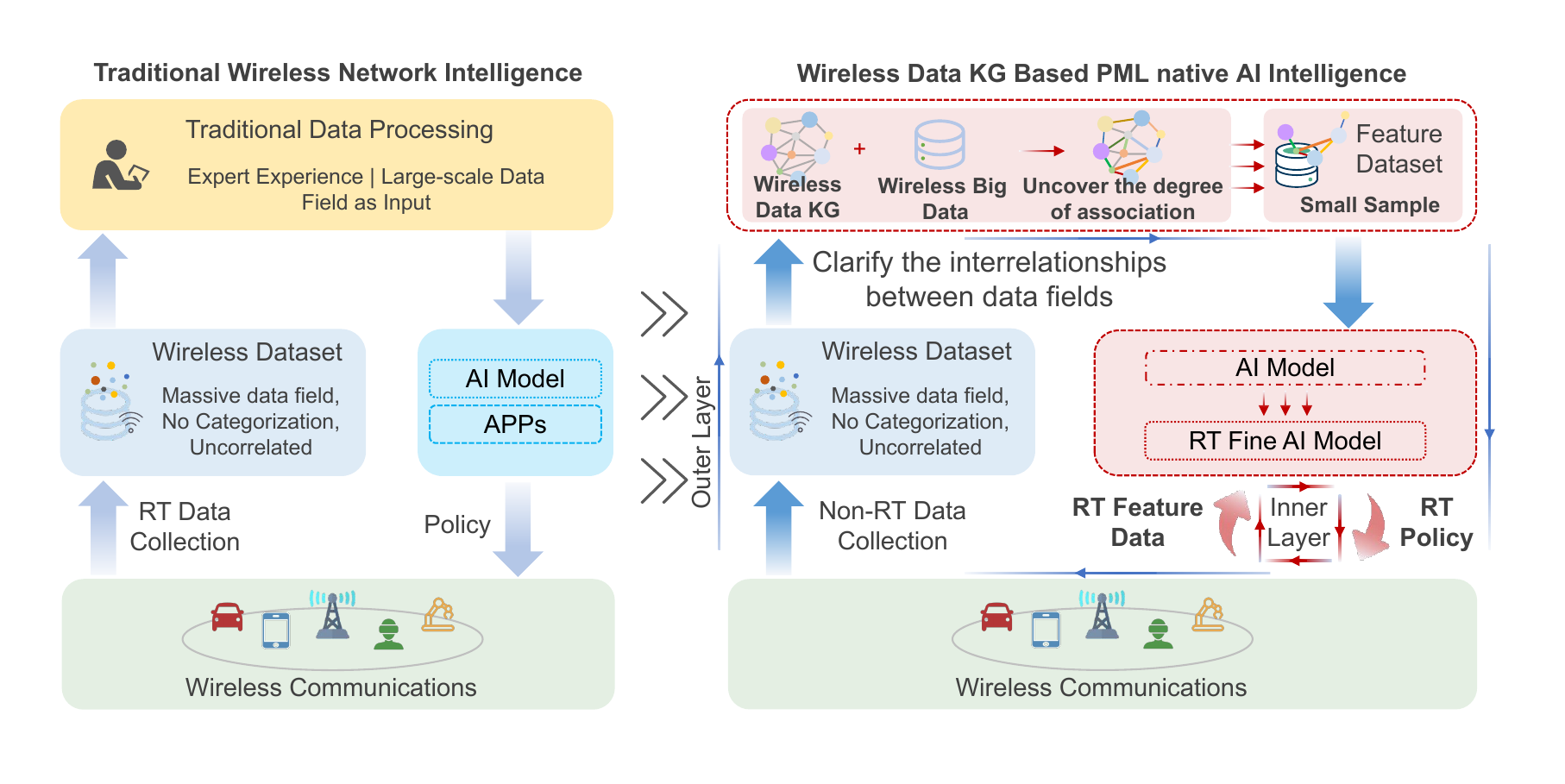}
	\caption{Comparison of two different wireless network intelligence frameworks}
	\label{intelligence_framework}
 \vspace*{-1.5em}
\end{figure}

In this section, we first propose a PML native AI architecture based on the wireless data KG, as shown on the right side of Fig. \ref{intelligence_framework}. In contrast, traditional wireless network intelligence is depicted on the left side of Fig. \ref{intelligence_framework}. The current wireless network intelligence primarily relies on real-time collection of wireless big data to drive AI models for intelligent network optimization. Due to the diversity of wireless data types, it typically requires high-dimensional datasets and large-scale AI networks. The data collection, AI training, and inference entail substantial costs, making it challenging to meet real-time and low-power requirements of wireless native AI. For the PML native AI architecture based on the wireless data KG depicted in Fig. \ref{intelligence_framework}, we propose for the first time the development of a wireless data KG to accurately utilize key features of wireless small data, enabling lightweight and green real-time native AI. The proposed architecture consists of a non-real-time outer layer and a real-time inner layer. In the outer layer, wireless big data is collected in a non-real-time manner. We semi-dynamically learn and construct a wireless data KG, analyzing, understanding, and representing the current intrinsic relationships between data fields. This allows us to identify a critical subset of feature data that influences the current KPI. Guided by the outer layer, the inner layer real-time collects a significantly reduced-scale feature data set and drives real-time AI training and inference, thereby achieving efficient real-time native AI for wireless networks. In the proposed PML native AI architecture, we only need to collect a small amount of key data fields in real time, thereby being able to train lightweight AI models, thus reducing the costs associated with data collection and computation, and supporting the realization of real-time, green network intelligence.

Furthermore, this section introduces the concept of the wireless data KG and provides a detailed description. Additionally, we offer an illustrative example of the wireless data KG, with a focus on throughput as a KPI.

\subsection{Definition and Characterization of Wireless Data KG}


In contrast to traditional KGs, the wireless data KG possesses several distinctive properties. Accordingly, the following offers a definition and a comprehensive characterization of the wireless data KG.

\begin{definition}\label{def1}
	\textbf{Wireless Data Knowledge Graph (wireless data KG)}
	is a KG that comprehensively portrays the association among the various factors in the environment, device properties, and the complete flowchart protocol stack within wireless communication networks. A wireless data KG can be denoted as $\mathcal{G} = \{\mathcal{V},\mathcal{E}, \mathbf{W}, \mathbf{T}, \mathbf{A}, \mathbf{X}\}$.
\end{definition}

The meanings of the symbols in the above definitions are described respectively below.

$\bullet$ $\mathcal{G}$ denotes a wireless data KG. For the sake of better examples in this paper, $\mathcal{G}$ can refer to the global wireless data KG expressed in Definition \ref{def1}, or to a wireless data KG that portrays a certain local environment of wireless communication with a KPI or several KPIs as core nodes.

$\bullet$ $\mathcal{V}$ denotes the set of all nodes in $\mathcal{G}$, where the $i$-th node is indicated by $v_i$ with the number of nodes $|\mathcal{V}| = N$. Each node in $\mathcal{V}$ corresponds to the various factors in Definition \ref{def1}, collectively referred to wireless data fields.


To distinguish the different types of nodes, $v_i$ is denoted by $(\mathbf{s})_i$, where $\mathbf{s} \in \mathbb{R}^N$ represents the node type vector of all nodes. Let $\Phi : \mathcal{V} \rightarrow \mathcal{S}$ be the node type mapping function, where $\mathcal{S}$ denotes the set of node types.

$\bullet$ $\mathcal{E}$ is the set of all edges in $\mathcal{G}$, $e_{i,j}$ indicates the connections between $v_i$ and $v_j$. Guided by wireless protocols and communication principles, the correlations between wireless data fields are determined. However, in real communication scenarios, these correlations between wireless data fields may not always be established. In other words, the edges between these nodes may change over time.


There are also multiple types of edge $e_{i,j}$, which can be denoted by relation type $(\mathbf{R})_{i,j}$, where $\mathbf{R} \in \mathbb{R}^{N \times N}$ indicates the relation type matrix. Let $\Psi : \mathcal{E} \rightarrow \mathcal{R}$ be the relation type mapping function, where $\mathcal{R}$ denotes the set of relation types.

$\bullet$ $\mathbf{W}\in \mathbb{R}^{N\times F}$ is a static attribute matrix representing the fixed attributes associated with each node. Each row of $\mathbf{W}$ denotes a node and columns indicate $F$ attributes. The fixed attributes are determined according to the protocol, such as node type, communication layer and adjustability.


$\bullet$ $\mathbf{T}=\{t_1,t_2,\cdots\}$ primarily reflects the temporal nature of the wireless data KG, where $0<t_1<t_2<\cdots$ and $t_i\in\{t_1,t_2,\cdots\}$ is a sampling time. Furthermore, $t_i$ and $t_{i+1}$ represent adjacent sampling times, and $t_1,t_2,$ and all subsequent $t_i$ add up to a contiguous sampling time period. The importance of $\mathbf{T}$ is emphasised because the wireless data KG may have different graph structures at these sampling moments, i.e., the wireless data KG is a continuous time dynamic graph. The wireless data KG is modeled as a sequence of time-stamped events $\mathcal{G}=\{G(t_1),G(t_2),\cdots\}$, representing the graph structure corresponding to each instance of communication, which may be the same or different, at each sampling time.

\begin{figure}[!t]
    \centering
    \includegraphics[width=4.1in]{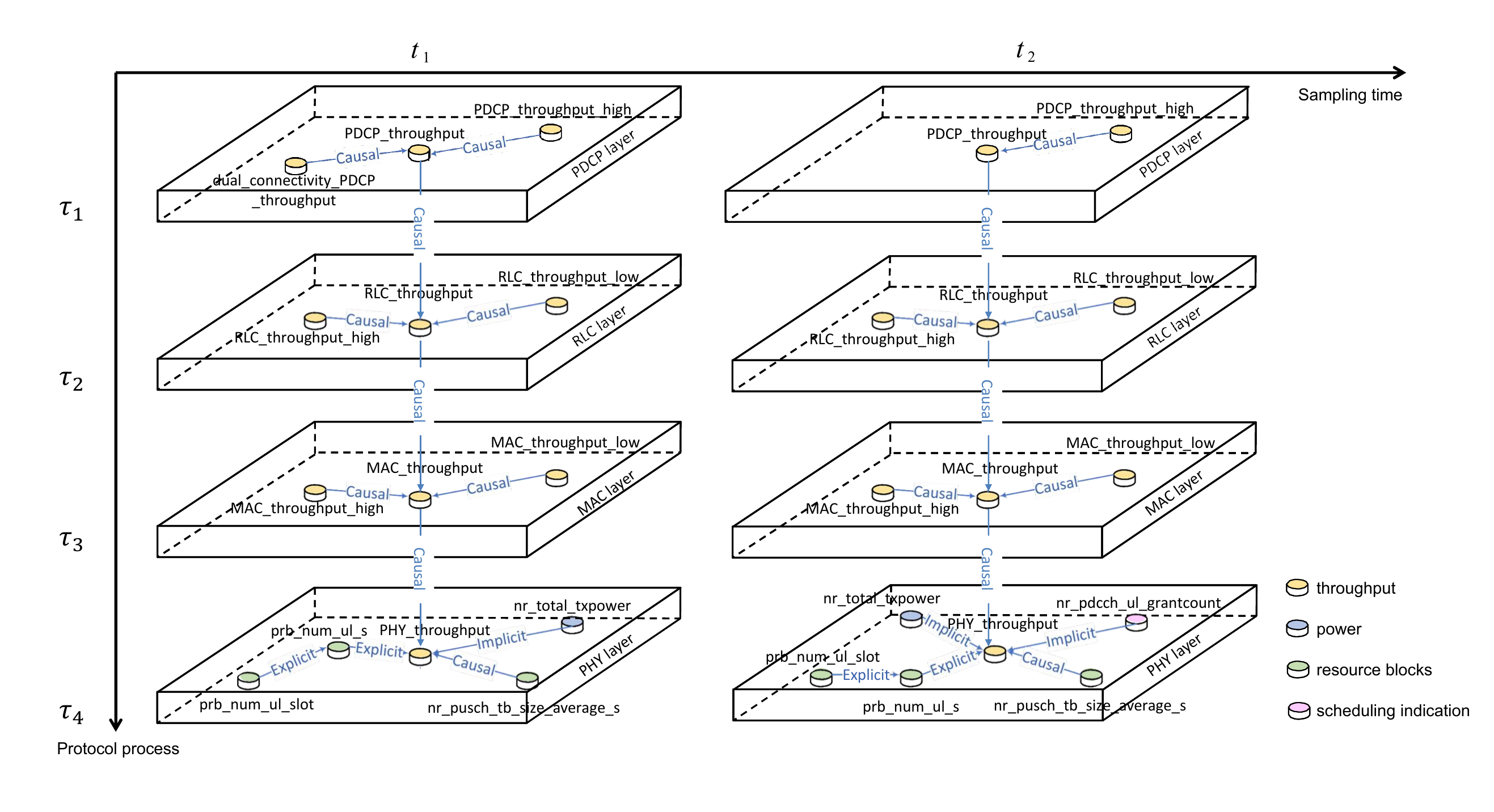} 
    \caption{Dynamic graph model of wireless data KG.}
    \label{time_series}
\end{figure}

Actually, a wireless data KG has two timelines: protocol process and sample time series as shown in Fig. \ref{time_series}. Starting with the protocol process, we have decided to use the Service Data Adaptation Protocol (SDAP) layer, Packet Data Convergence Protocol (PDCP) layer, Radio Link Control (RLC) layer, Medium Access Control (MAC) layer, and Physical (PHY) layer for the wireless access network. The influence between nodes within the same layer is considered simultaneous, while the influence between different layers follows the chronological order according to the protocol. For instance, in the uplink, MAC layer throughput determined at $\tau_3$ can have impact on the subsequent PHY layer throughput at $\tau_4$. Nevertheless, the time difference introduced by the protocol process is negligible, allowing the different layers to be treated as the same timestamp in the sampling timeline. The layer to which these nodes belong is also one of their attributes, and the rest attributes such as node type will be described in the following. Unlike a static graph where relations remain constant, the graph structure keeps evolving during the sampling process in the wireless data KG. For example, at sample time $t_1$, the \texttt{dual\_connectivity\_PDCP\_throughput} has an effect on the \texttt{PDCP\_throughput}. However, due to the changes in channel state and communication tasks, this effect may dissipate at sample time $t_2$.

\begin{figure}[!t]
	\centering
	\includegraphics[width=3.6in]{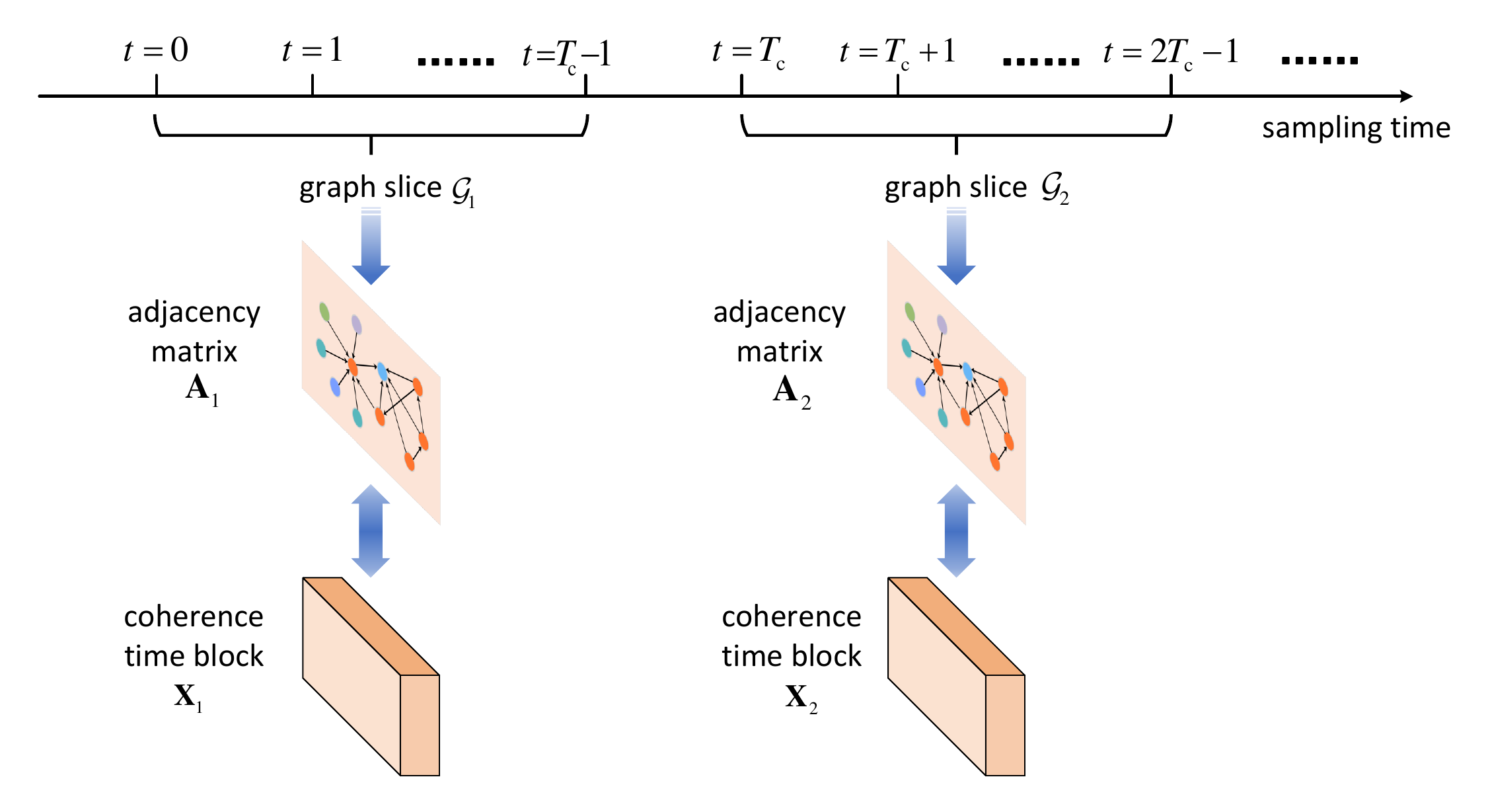}
	\caption{Illustration of graph slice model.}
	\label{graph_slice}
\end{figure}

From the above analysis, we readily see that the topology of the wireless data KG changes over time, although not continuously. In particular, the wireless communication network channel state is stable within each the coherence time. Therefore, the wireless data KG topology can be determined with the assistance of coherence time. In our scenario, where a moving car consistently sends and receives signals around several base stations, the coherence time is computed by
\begin{spacing}{0.6}
\begin{equation}
	T_\mathrm{c} = \frac{1}{f_\mathrm{m}} = \frac{\lambda}{v \cos{\theta}},
\end{equation}
\end{spacing}
\noindent where $T_\mathrm{c}$ and $f_\mathrm{m}$ denote the coherence time and Doppler shift, respectively, and $\lambda$, $v$, and $\theta$ are the wavelength, car movement speed and clip angle, respectively. Upon the determination of coherence time $T_\mathrm{c}$, the wireless data KG can be segmented into discrete graph slices as shown in Fig. \ref{graph_slice}. The coherence time switch point is $mT_\mathrm{c}$, where $m\in \mathbb{N^+}$. In other words, the graph from ${(m-1)T_\mathrm{c}}$ to ${mT_\mathrm{c}-1}$ share the same topology, determined by the aforementioned construction process. The $m$-th graph slice is denoted as $\mathcal{G}_m$ and the total number of graph slices is $M$.

\begin{figure}[!t]
	\centering
	\includegraphics[width=3.6in]{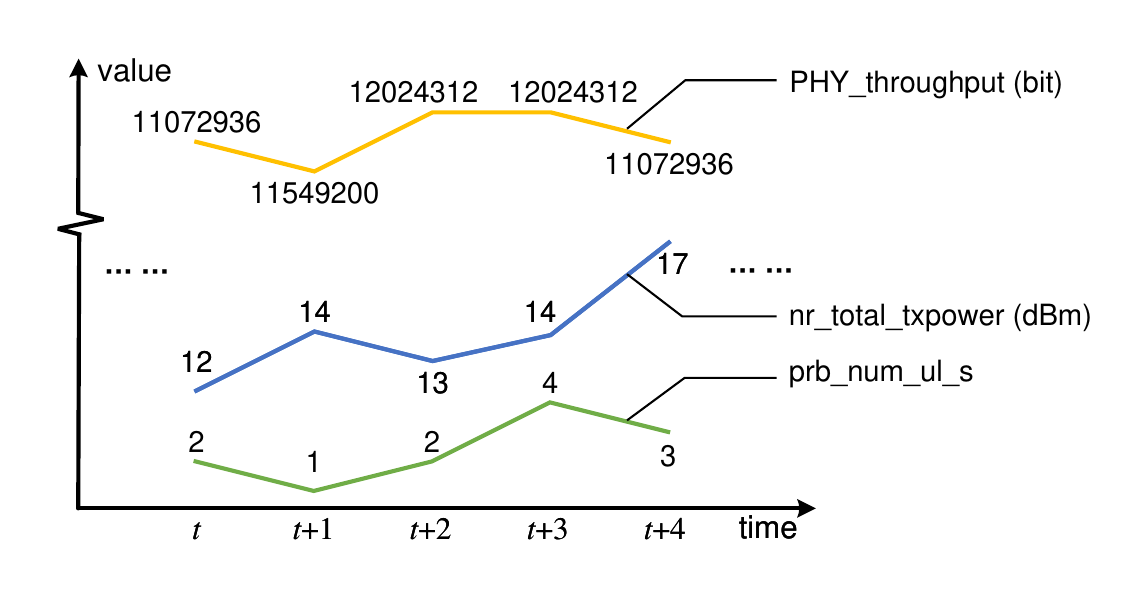}
    \vspace*{-1.5em}
	\caption{Wireless big data collected by each node.}
	\label{data}
    \vspace*{-1.5em}
\end{figure}

Therefore, the wireless data KG during a sampling time period $T$ can be modeled as a series of graph slices $\{\mathcal{G}_1, \mathcal{G}_2, \ldots, \mathcal{G}_M\}$. The graph slice $\mathcal{G}_i$ represents numerous sampling instances, each corresponding to a $G(t_j)$, signifying that the graph structure of these $G(t_j)$ remains unchanged within $\mathcal{G}_i$. The $m$-th graph slice can be denoted by $\mathcal{G}_m = \{\mathcal{V},\mathcal{E}_m, \mathbf{W}, \mathbf{A}_m, \mathbf{X}_m\}$, where $\mathcal{V}$ and $\mathbf{W}$ remain constant; in other words, the number of nodes and node attributes in the wireless data KG stays consistent over time. However, $\mathcal{E}_m$, $\mathbf{A}_m$, and $\mathbf{X}_m$ vary elaborated in the following.

$\bullet$ $\mathbf{A}$ denotes the adjacency matrix corresponding to the wireless data KG at each moment $t$, where $t\in T$. The element $\mathbf{A}_{i,j}$ in $i$-th row and $j$-th column indicates whether $v_i$ and $v_j$ are connected, which is defined as
\begin{spacing}{0.6}
\begin{equation}
\mathbf{A} = (\mathbf{A})_{i,j} \in \mathbb{R}^{N\times N}, ~~(\mathbf{A})_{i,j} =
\begin{cases}
1, & \text{if } (v_i, v_j) \in \mathcal{E} \\
0, & \text{otherwise.}
\end{cases}
\end{equation}
\end{spacing}
\noindent and $\mathbf{A}_m$ represents the adjacency matrix of the graph slice. Since the graph structure of wireless data KG changes over time, $\mathbf{A}_j$ varies with $\mathcal{G}_i$.

$\bullet$ $\mathbf{X}$ denotes the matrix formed by the real wireless data collected by each node in the wireless data KG. Within a coherence time period, the wireless big data can be collected at each time $t$. In Fig. \ref{data}, the data formats of the three selected entities are presented to demonstrate this feature. It worth to note that this authentic data is generated from the true-data testbed for 5G/B5G intelligent network (TTIN), which is the first real-world platform for real-time wireless data collection, storage, analytics, and intelligent closed-loop control \cite{TTIN}. Let the real data collected of node $v_i$ at time $t$ be $x^i_t \in \mathbb{R}$. Hence, the data collected by all $N$ nodes at time $t$ can form a data vector $\mathbf{x}_t = [x^1_t, x^2_t, \ldots, x^{N}_t]^\mathsf{T} \in \mathbb{R}^{N}$. Accordingly, the data matrix corresponds to the graph slice $G_m$ is written as $\mathbf{X}_m$, which consists of a series of data vectors $\mathbf{X}_m =  \left[\mathbf{x}_{(m-1)T_\mathrm{c}}, \mathbf{x}_{(m-1)T_\mathrm{c}+1}, \ldots, \mathbf{x}_{mT_\mathrm{c}-1} \right] \in \mathbb{R}^{N \times T_\mathrm{c}}$.

\subsection{Exploring Wireless Data KG: An Illustrative Example}

\begin{figure}[!t]
	\centering
	\includegraphics[width=3.5in]{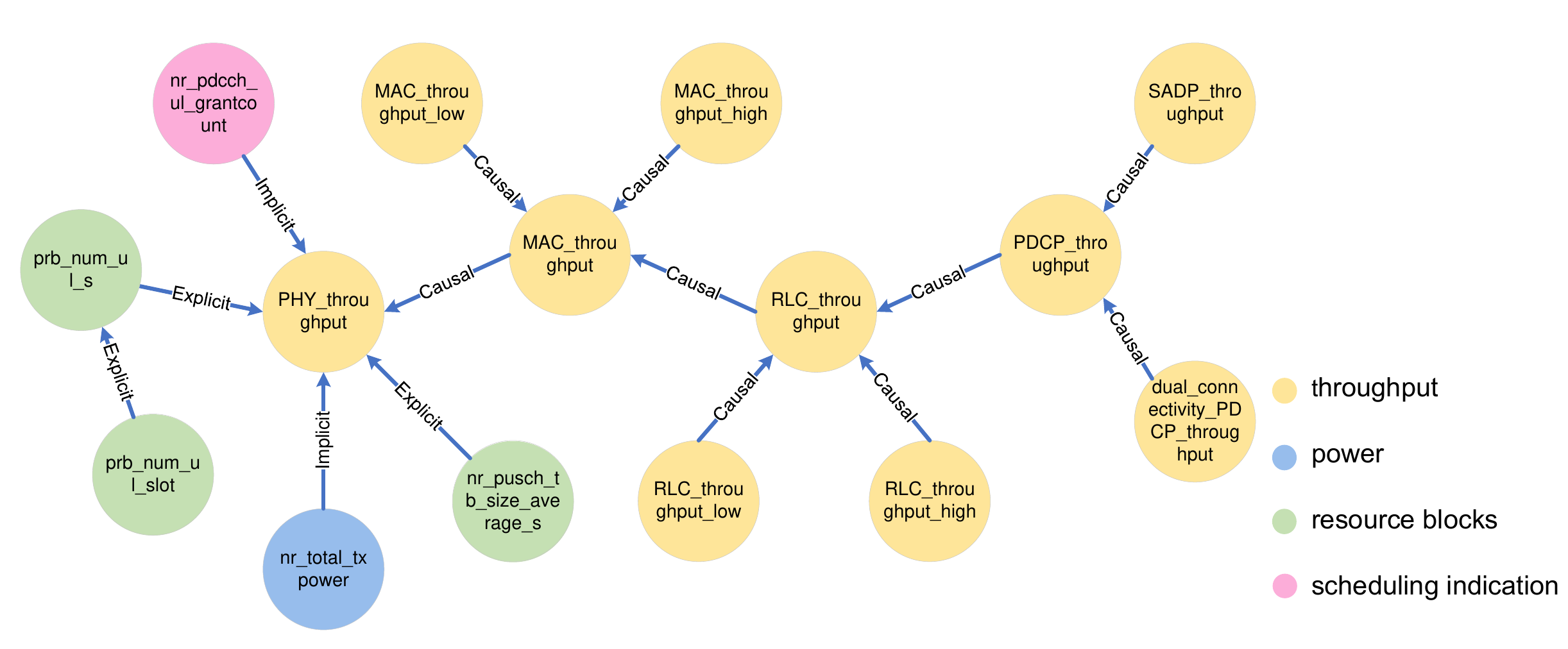}
	\caption{A partial of the constructed uplink throughput wireless data KG.}
	\label{wireless data KG_example}
    \vspace*{-1.5em}
\end{figure}

In this section, we provide an example of a wireless data KG, based on the technical specification 21.205 of the 3GPP Release 17 \cite{3GPP}. According to the aforementioned definition, a wireless data KG visually and in real-time depicts the correlation between different wireless data fields. In the practical construction of the wireless data KG, an illustrative example corresponding to a graph slice within a coherence time is presented here to offer a concise and clear representation. The subsequent paragraphs will use the uplink throughput wireless data KG fragment as an example to intuitively showcase the fundamental elements of the wireless data KG. A segment of the constructed uplink throughput wireless data KG is visualized in Fig. \ref{wireless data KG_example}.

\begin{table}[ht]
\begin{spacing}{1.5}
\fontsize{6pt}{5pt}\selectfont
	\caption{Edge classification in wireless data KG \label{tab:edges}}
	\centering
	\begin{tabular}{ccp{3.3cm}p{2.5cm}}
		\hline
		Category & Number & \makecell[c]{Definition} & \makecell[c]{Example} \\
		\hline
		Causal Relation & 70 & \makecell[c]{Causal relation indicates a strong link between\\ two entities with a direct causal influence.} & \makecell[c]{\texttt{MAC\_throughput} \& \\ \texttt{PHY\_throughput} } \\
		Explicit Relation & 35 & \makecell[c]{Explicit relation describes a less tight link \\ with specific expression.} & \makecell[c]{\texttt{prb\_num\_ul\_s} \& \\ \texttt{PHY\_throughput} }\\
		Implicit Relation & 28 & \makecell[c]{Implicit relation describes a less tight link \\ without specific expression.} & \makecell[c]{\texttt{nr\_total\_txpower} \& \\ \texttt{PHY\_throughput} } \\
		\hline
		\textbf{Total} & \textbf{133} & \makecell[c]{\textbf{/}} & \makecell[c]{\textbf{/}} \\
		\hline
	\end{tabular}
    \normalsize
\end{spacing}
 \vspace*{-1.5em}
\end{table}

An uplink throughput wireless data KG centers around the uplink throughput and visually represents the relationships among 82 nodes in the form of a graph. Figure \ref{wireless data KG_example} depicts a local view of the uplink throughput wireless data KG. In this representation, nodes of different colors represent different types of entities, categorized into nine classes based on their physical attributes, namely: 1) throughput, 2) power, 3) scheduling indication, 4) modulation encoding indication, 5) resource blocks, 6) block error rate, 7) switch indication, 8) antenna configuration indication, and 9) frame structure. Thus, there are a total of 9 categories denoted by symbol $\mathcal{S}$. Each pair of interconnected nodes signifies a relationship between them, categorized into three types: causal relation, implicit relation, and explicit relation, i.e., $\mathcal{R} = \{\mathrm{causal}, \mathrm{implicit}, \mathrm{explicit}\}$. A total of 133 relations are identified in the uplink throughput wireless data KG, and the relation between any two entities belongs to $\mathcal{R}$. The definitions and examples of these three types of relationships can be referred to in Table  \ref{tab:edges}.

\section{Construction and Application of Wireless Data KG}

This section primarily delves into the pathways to achieve PML native AI, with a focus on exploring the wireless data KG. The first task is to construct a wireless data KG by integrating knowledge and data. The second task involves generating a feature dataset based on the constructed wireless data KG. A brief description of the implementation process and technical approach for these two tasks is provided, laying the groundwork for the subsequent specific algorithm designs.

\subsection{Construction of Wireless Data KG with a Blend of Knowledge and Data}

Acknowledging the dynamic nature of the constructed wireless data KG, with complex relationships evolving over time, manual construction poses challenges due to significant labor costs and time overheads. The inherent subjectivity in human decision-making introduces the possibility of errors and omissions during the construction process. Therefore, a more desirable approach involves the synergistic integration of both knowledge and data to streamline the wireless data KG construction. This strategic combination harnesses the insights gleaned from manually constructed local wireless data KGs and tapping into the vast potential of wireless big data. By doing so, the generation and refinement of the remaining portions of the wireless data KG can be achieved with greater efficiency. This integrated approach not only enhances accuracy but also contributes to a notable reduction in construction costs.

This subsection delineates an intelligent approach to construct a wireless data KG by strategically leveraging expert/protocol knowledge in conjunction with wireless big data. Importantly, this approach avoids the need for specific experimental processes. The emphasis here is on explaining the processes of graph embedding learning and graph link prediction tailored for the wireless data KG.

\subsubsection{Wireless data KG graph embedding formulation}

With multiple sources of information given, useful information can be extracted and the high-dimensional raw data can be compressed into a low-dimensional representation vector, thereby facilitating subsequent manipulation.  This boils down to a graph embedding problem.
\begin{definition}
	\textbf{Graph Embedding.}
	Given a graph $\mathcal{G} = \{\mathcal{V},\mathcal{E}, \mathbf{W}, \mathbf{T}, \mathbf{A}, \mathbf{X}\}$, graph embedding is the task to learn the $c$-dimensional embedding matrix $\mathbf{Z} \in \mathbb{R}^{N\times c}$ for all $v_i \in \mathcal{V}$ that are able to capture the rich structural and semantic information.
\end{definition}

Graph embedding for a wireless data KG poses several challenges. Foremost among these challenges is the tremendous amount of wireless data collected by the nodes in the graph. This data holds vast potential information, intensifying the complexity of its embedding. To tackle this, wireless big data is processed in batches corresponding to the graph slices and undergoes subsequent processing with a graph neural network (GNN) post time convolution processing.


Secondly, the wireless data KG is characterized by its composition as an attribute graph, incorporating various types of nodes and edges, thus exhibiting heterogeneity \cite{Wang2019Ji}. This makes it challenging to mine nodes and edges for multiple attributes. To address this challenge, a concept of meta-path is introduced. Then, the previously mentioned GNN will be transformed into a heterogeneous graph attention neural network. This section will elaborate on the utilization of these heterogeneities by meta-paths. In the wireless data KG, various relation types encapsulate distinct semantic information, signifying different degrees of influence. Consequently, the significance of relation types surpasses that of node types, thereby introducing the notion of generalized meta-paths.

\begin{figure}[!t]
	\centering
	\includegraphics[width=3.5in]{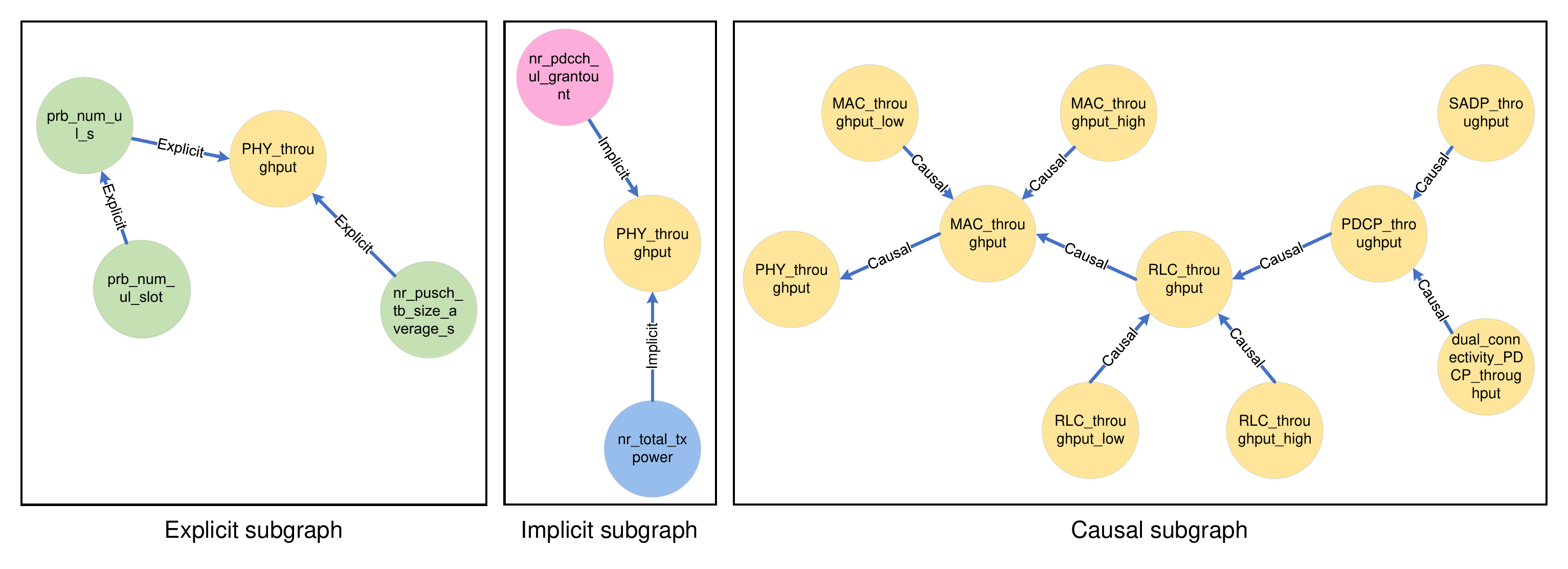}
	\caption{Illustration of explicit, implicit, and causal subgraphs.}
	\label{meta_path}
    \vspace*{-1.5em}
\end{figure}

\begin{definition}
	\textbf{Generalized Meta-path.}
	A generalized meta-path $\phi_p$ is defined as a path in the form of $\cdot \stackrel{R_1}{\longrightarrow} \cdot \stackrel{R_2}{\longrightarrow} \cdots \stackrel{R_l}{\longrightarrow} \cdot$ (abbreviated as $R_1R_2...R_{l}$, where $\left(\cdot\right)$ denotes a node of any type), which describes a composite relation $R = R_1 \circ R_2 \circ \cdots \circ R_l$ between nodes, where $\circ $ denotes the composition operator on relations.
\end{definition}
\textit{Example.} As shown in Fig. \ref{meta_path}, three generalized meta-paths, $\cdot \stackrel{\textrm{causal}}{\longrightarrow} \cdot$, $\cdot \stackrel{\textrm{implicit}}{\longrightarrow} \cdot$ and $\cdot \stackrel{\textrm{explicit}}{\longrightarrow} \cdot$, are defined, respectively. Accordingly, wireless data KG can be divided into three subgraphs, i.e., causal, implicit and explicit subgraphs. Different from the original meta-path definition, generalized meta-path only focuses on relation types rather than node and relation types. In what follows, generalized meta-path is simplified as meta-path.

Given a meta-path $\phi_p$, there exists a set of meta-path based neighbors of each node which can reveal diverse structure information and rich semantics in a heterogeneous graph.

\begin{definition}
	\textbf{Meta-path-based Neighbor.}
	Given a meta-path $\phi_p$ in a heterogeneous graph, the meta-path-based neighbors $\mathcal{N}^{\phi_p}_i$ of node $i$ are defined as the set of nodes that connect with node $i$ via meta-path $\phi_p$. Note that the node's neighbor $\mathcal{N}^{\phi_p}_i$ includes itself if $\phi_p$ is symmetric.
\end{definition}
\textit{Example.} Taking Fig. \ref{meta_path} as an example, given the explicit subgraph, the meta-path based neighbors of \texttt{PHY\_throughput} includes itself, \texttt{prb\_num\_ul\_s} and \texttt{nr\_pusch\_tb\_size\_average\_s}. Obviously, meta-path based neighbors can exploit different aspects of the structure information in a heterogeneous graph.

Thirdly, the wireless data KG is dynamic, which makes it harder to represent the continuous embedding of the evolving graph. In this regard, different from the static graph, a continuous dynamic graph embedding problem must be formulated. The objective is to devise a neural network model that can generate $c$-dimensional embedding for each graph slice. Specifically, given a series of graph slices $\{\mathcal{G}_0, \mathcal{G}_1, \dots, \mathcal{G}_M\}$, a series of embedding matrix need to be generated for each graph slice. That is
\begin{spacing}{0.7}
\begin{equation}
	\{ \mathbf{Z}_0, \mathbf{Z}_1, \ldots, \mathbf{Z}_M \}= f(\mathcal{G}_0, \mathcal{G}_1,
	\ldots, \mathcal{G}_M),
\end{equation}
\end{spacing}
\noindent where $\mathbf{Z}_m = [\mathbf{z}^1_m, \mathbf{z}^2_m, \ldots, \mathbf{z}^N_m]^\mathsf{T}$ represents the embedding matrix for graph slice $\mathcal{G}_m$, and $\mathbf{z}_m^i$ indicates the embedding vector of node $v_i$ in graph slice $\mathcal{G}_m$. Then, downstream applications such as link prediction can be performed based on the obtained embedding vectors.

\subsubsection{Wireless data KG graph link prediction task}

Following the acquisition of node embedding representation vectors for the wireless data KG in the preceding section, the subsequent phase involves employing a similarity function. This function converts the vectors of two nodes into a measure of the degree of association between them. This measure of relational association  can be subsequently utilized to ascertain whether a connection exists between the nodes, which aligns with the objective of graph link prediction.

\subsection{Intelligent Generation of Feature Dataset Based on Wireless Data KG}

The main purpose of this section is to verify that our proposed PML native AI architecture can achieve green and lightweight intelligence. The work primarily involves the generation of feature datasets and the evaluation of these datasets.

\subsubsection{Feature selection based on wireless data KG}

In order to identify a subset of critical nodes from a large volume of wireless data fields that have the most substantial impact on the target KPI, we leverage a wireless data KG for feature selection. Here, each node represents a feature related to the KPI node. Initially, the graph structure is used to identify all paths connecting the KPI. Subsequently, the influence of each node on the KPI is determined based on the relationship between neighboring nodes on the paths. The degree of relationship between neighboring nodes can be measured using node similarity in link prediction tasks. The nodes are sorted according to their impact on the KPI. Finally, feature ranking is employed to guide the selection of features.

\subsubsection{Feature dataset generation and evaluation}

After selecting important nodes based on their impact on the KPI from a plethora of wireless data fields, we proceed to evaluate the feature dataset to ensure that we have identified a minimally sized subset that maximizes information content and importance. Two metrics are utilized for the assessment and optimization of the feature dataset. The first metric is the goodness of fit, which involves utilizing the selected nodes and the corresponding collected data to predict the target KPI and calculating the disparity between predicted values and actual values. In practice, the goodness of fit needs to be ensured at a certain level based on real-world scenarios. The second metric is the feature compression ratio. Given the prerequisite of ensuring a good fit, feature selection is performed based on the feature compression ratio to maximize the retention of the most essential information within the selected features and to minimize redundancy. This approach reduces costs and aligns with the requirements of green intelligence.

\section{Methodology for Construction and Application of Wireless Data KG}

In this section, we present two specific algorithms for constructing and applying wireless data KGs. The first algorithm is the STREAM framework, designed for constructing the wireless data KG, while the second algorithm is the feature dataset generation algorithm based on the wireless data KG. This section offers a detailed description of the implementation process of these two algorithms.


\subsection{Wireless Data KG Graph Embedding}

\begin{figure}[!t]
	\centering
	\includegraphics[width=3.5in]{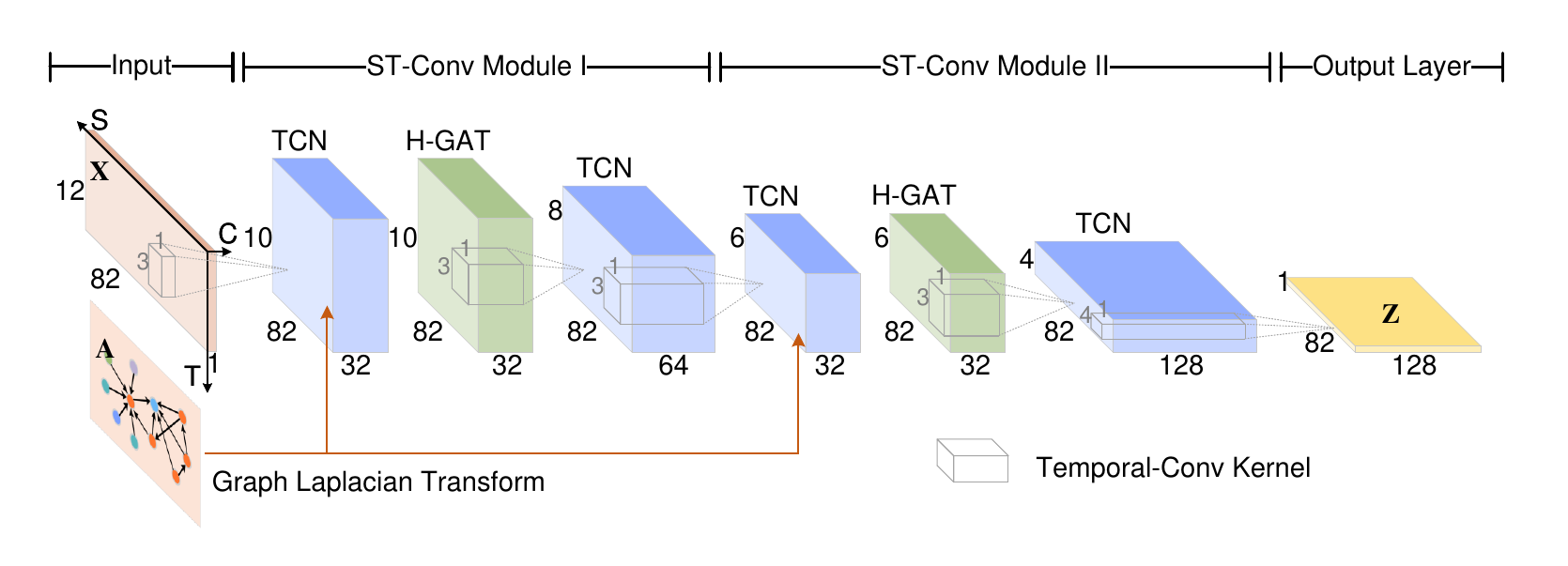}
	\caption{Model framework dimension analysis.}
	\label{model_framework}
 \vspace*{-1.5em}
\end{figure}

In this section, a general framework tailored for the intelligent construction of wireless data KG is described, taking into account the salient features of wireless data KG as well as wireless big data. The proposed framework, named STREAM, employs the spatial-temporal graph neural network to leverage information from topology, data matrix, and node attributes. It incorporates a hierarchical attention mechanism to handle the heterogeneity of nodes and edges. The overall framework, illustrated in Fig. \ref{model_framework}, consists of an input layer, two stacked spatial-temporal convolution (ST-Conv) modules, and an output layer. Each ST-Conv module comprises two temporal convolutional layers and one spatial convolutional layer, which effectively exploits the spatio-temporal nature of wireless data KG. Moreover, the spatial convolution layer adopts a hierarchical attention mechanism, i.e., node-level aggregation is performed in each subgraph firstly, and then meta-path-level aggregation is carried out for the entire graph. We refer to this layer as the heterogeneous graph attention network, abbreviated as H-GAT. Details of these convolution layers are explained in Fig \ref{model_framework}.

\begin{figure}[!t]
	\centering
	\includegraphics[width=3.5in]{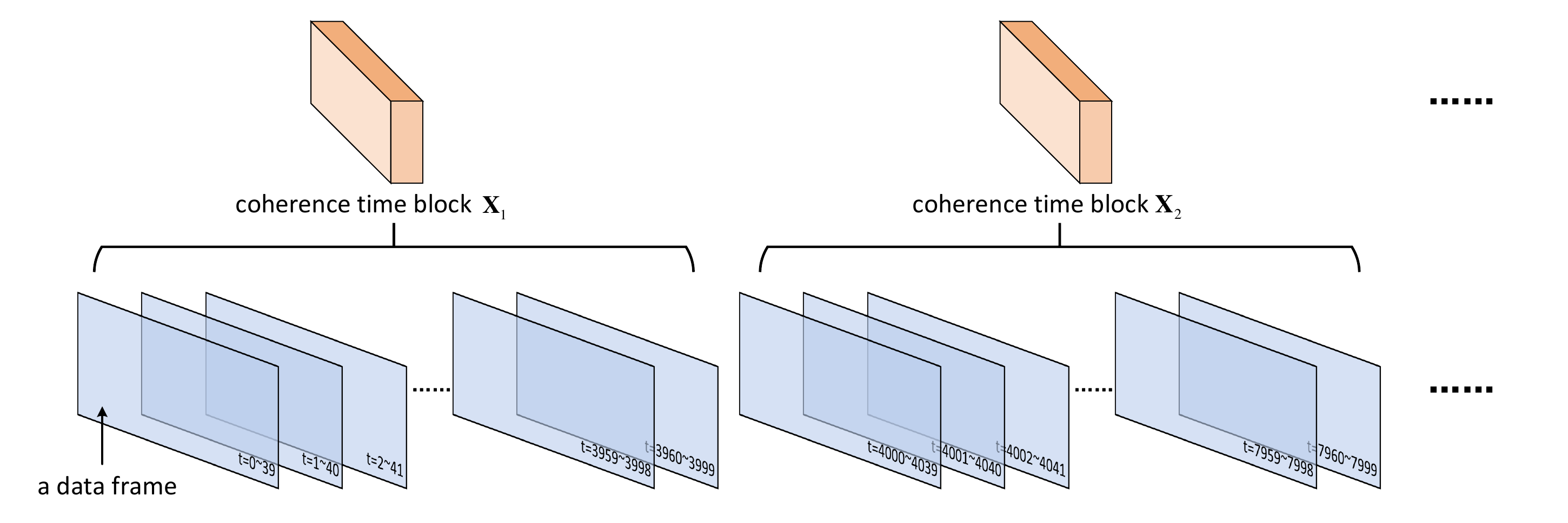}
	\caption{Illustration of data frames.}
	\label{data_frame}
\vspace*{-1.5em}
\end{figure}

To tackle the issue arising from the extended data length of the coherence time block, hindering its direct involvement in temporal convolution, a crafted data segmentation approach is presented, depicted in Fig. \ref{data_frame}. To uphold time dependency, a coherence time block is partitioned into multiple overlapping data frames. No overlap is permitted between different coherence time blocks. It is noteworthy that the length of the data frame can be adaptively adjusted. Extremely short data frames are ineffective in capturing time dependencies, whereas excessively long frames can increase the computational burden.
\begin{spacing}{1}
\end{spacing}
\subsubsection{Spatial convolution layer}
Graph data is a typical non-Euclidean data and cannot be processed by standard convolution operation, so we employ the graph convolution.
Firstly, the adjacency matrix of the $m$-th graph slice $\mathbf{A}_m$ is simplified as $\mathbf{A}$, and the normalized adjacency matrix $\widetilde{\mathbf{A}}$ is defined by
\begin{spacing}{0.7}
\begin{equation}
	\widetilde{\mathbf{A}} = \mathbf{A} + \mathbf{I},
\end{equation}
\end{spacing}
\noindent where $\mathbf{I} \in \mathbb{R}^{N\times N}$ is the identity matrix.
The degree matrix is  defined as a diagonal matrix $\mathbf{D}$  with $(\mathbf{D})_{i,i} = \sum_{j=1}^N (\mathbf{A})_{i,j}$. Similarly, the normalized degree matrix is defined as $\widetilde{\mathbf{D}}$ with its diagonal element $(\widetilde{\mathbf{D}})_{i,i} = \sum_{j=1}^N (\widetilde{\mathbf{A}})_{i,j}$.
Let $\boldsymbol{\Theta}$ be a graph convolution kernel. Combining with the activation function $\sigma$, the multilayer propagation rule of GCN can be written as,
\begin{spacing}{0.7}
\begin{equation}
	\mathbf{H}^{l+1} = \sigma\left( \boldsymbol{\Theta} \circledcirc \mathbf{H}^l\right) =\sigma\left({\widetilde {\bf{D}}^{ - 1/2}}\widetilde {\mathbf{A}}{\widetilde {\mathbf{D}}^{ - 1/2}}{\mathbf{H}^l}\mathbf{O}^l \right),
\end{equation}
\end{spacing}
\noindent where $\mathbf{O}$ is a trainable parameter matrix.
In the middle layers of the STREAM framework, the representation matrix becomes the representation tensor due to existence of multiple channels. Therefore, the graph convolution needs to be generalized to 3-dimensional, the convolution result of the $j$-th kernel can be calculated as follows,
\begin{spacing}{0.7}
\begin{equation}
	\left(\mathcal{H}\right)^{l+1}_j = \sum^{c_\mathrm{in}}_{i=1} \sigma\left( \widetilde{\mathbf{D}}^{-1/2}\widetilde{\mathbf{A}}\widetilde{\mathbf{D}}^{-1/2}\left(\mathcal{H}\right)^l_i \mathbf{O}^l_i \right), \quad 1\leq j \leq c_\mathrm{out},
	\label{gcn}
\end{equation}
\end{spacing}
\noindent where $c_\mathrm{in}$ and $c_\mathrm{out}$ indicate the input channel and output channel, respectively. Namely, a total of $c_\mathrm{out}$ kernel participate in the graph convolution of the $l$-th layer.
Particularly, the representation tensor of the first layer equals to the data matrix, i.e., $\mathcal{H}^0 = \mathbf{X}$.


$\bullet$ \emph{Node-level Attention Mechanism}

For a given meta-path, each node's neighbors play different roles in the graph embedding for a particular task and show different importance. Therefore, introducing node-level attention can learn the importance of meta-path-based neighbors for each node in aggregation.

For node pair $(v_i, v_j)$ on a given meta-path $\phi_p$, the node-level attention coefficients $ s^{\phi_p}_{i,j} $ of node $ i $ to node $ j $ are related to their own characteristics and can be calculated by
\begin{spacing}{0.7}
\begin{equation}
	s_{i,j}^{l, \phi_p}  = \sigma (({\mathbf{a}}^{\phi_p}) ^\mathsf{T} \cdot [{\mathbf{H}}_i^l\Vert{\mathbf{H}}_j^l]),
	\label{omega}
\end{equation}
\end{spacing}
\noindent where $\Vert$ denotes the vector concatenation operation, $\mathbf{H}^l_i$ denotes the embedding matrix of node $ i $ at the $ l $-th spatial convolutional layer, and $\mathbf{a}^{\phi_p}$ represents the node-level attention vector for meta-path $\phi_p$. After obtaining the attention coefficients based on meta-paths, they are normalized by the softmax function to obtain normalized attention coefficient $\widetilde{s}_{i,j}^{l,\phi_p} $:
\begin{spacing}{0.7}
\begin{equation}
	\widetilde{s}_{i,j}^{l, \phi_p} = \frac{\exp{(s^{l,\phi_p}_{i,j})}}{\sum_{k\in\mathcal{N}_i^{\phi_p}}\exp{(s^{l,\phi_p}_{i,k})}}.
	\label{tilde_omega}
\end{equation}
\end{spacing}
\noindent The obtained normalized node-level attention weight coefficients $\widetilde{s}_{i,j}^{l,\phi_p}$ can thus form a node-level coefficient matrix $\mathbf{S}^{l,\phi_p}$, where $(\mathbf{S})^{l,\phi_p}_{i,j}=\widetilde{s}_{i,j}^{l,\phi_p}$. Accordingly, the node-level coefficient matrix $\mathbf{S}^{l,\phi_p}$ can be directly multiplied with the embedding tensor $\mathcal{H}^l$:
\begin{spacing}{0.7}
\begin{equation}
	\mathcal{H}^{l, \phi_p} = \mathbf{S}^{l,\phi_p}\cdot \mathcal{H}^l,
	\label{node_level_agg}
\end{equation}
\end{spacing}
\noindent where $\mathcal{H}^{l, \phi_p}$ is the learned embedding tensor for meta-path $\phi_p$. The embedding of each node is obtained by performing the aggregation on its neighbors. Furthermore, given a set of meta-paths $\{\phi_1, \phi_2,...,\phi_P\}$, we can obtain $P$-group specific semantic embedding tensors, denoted by $ \{\mathcal{H}^{l, \phi_1}, \mathcal{H}^{l, \phi_2}, ..., \mathcal{H}^{l, \phi_P}\}$.



$\bullet$ \emph{Meta-path-level Attention Mechanism}

In general, each node in a heterogeneous graph contains multiple types of semantic information. Graph embedding based on a specific meta-path provides insight into only one facet of the node's semantics. To learn a more comprehensive graph embedding, the specific semantics embedded in each meta-path must be fused. To address this issue, we employ an meta-path-level attention mechanism. This mechanism automatically learns the importance of different meta-paths and fuse them to a specific task. Consequently, the importance of meta-path $\phi_p$, denoted by $e^{l,\phi_p}$, can be calculated by:
\begin{spacing}{0.7}
\begin{equation}
	{e^{l, {\phi _p}}} = \frac{1}{{|{\mathcal {V}}|}}\sum\limits_{i \in {\mathcal{ V}}} {{{\mathbf{r}}^\mathsf{T}} \cdot \tanh ({\mathbf{Q}} \cdot {\mathcal{H}}^{l,\phi_p}  + {\mathbf{b}})},
	\label{nu}
\end{equation}
\end{spacing}
\noindent where $\mathbf{Q}$ is the learnable parameter matrix, $\mathbf{b}$ is the bias, and $\mathbf{r}$ is the meta-path-level attention vector. After obtaining the importance of each meta-path, it is normalized by the softmax function. The normalized meta-path level attention coefficient of the meta-path $\phi_p$, denoted by $\widetilde{e}^{l,\phi_p}$, can be calculated by:
\begin{spacing}{0.7}
\begin{equation}
	\widetilde{e}^{l,\phi_p} =\frac{\exp(e^{l,\phi_p})}{\sum_{p=1}^P\exp(e^{l,\phi_p})} .
	\label{tilde_nu}
\end{equation}
\end{spacing}

This normalization can be interpreted as the contribution of meta-path $\phi_p$ to a particular task with the higher $\widetilde{e}^{l,\phi_p}$ is, the more important indicating greater importance for meta-path $\phi_p$. For different tasks, meta-path $\phi_p$ may have different weights. The learned weights serve as coefficients to merge these semantically specific embeddings, resulting in the final representation matrix of the $l$-th layer $\mathcal{H}^{l}$,
\begin{spacing}{0.7}
\begin{equation}
	 \mathcal{H}^{l}=\sum_{p=1}^P\widetilde{e}^{l,\phi_p}\cdot\mathcal{H}^{l, \phi_p} .
	 \label{meta_level_agg}
\end{equation}
\end{spacing}

\begin{figure}[!t]
	\centering
	\includegraphics[width=3.5in]{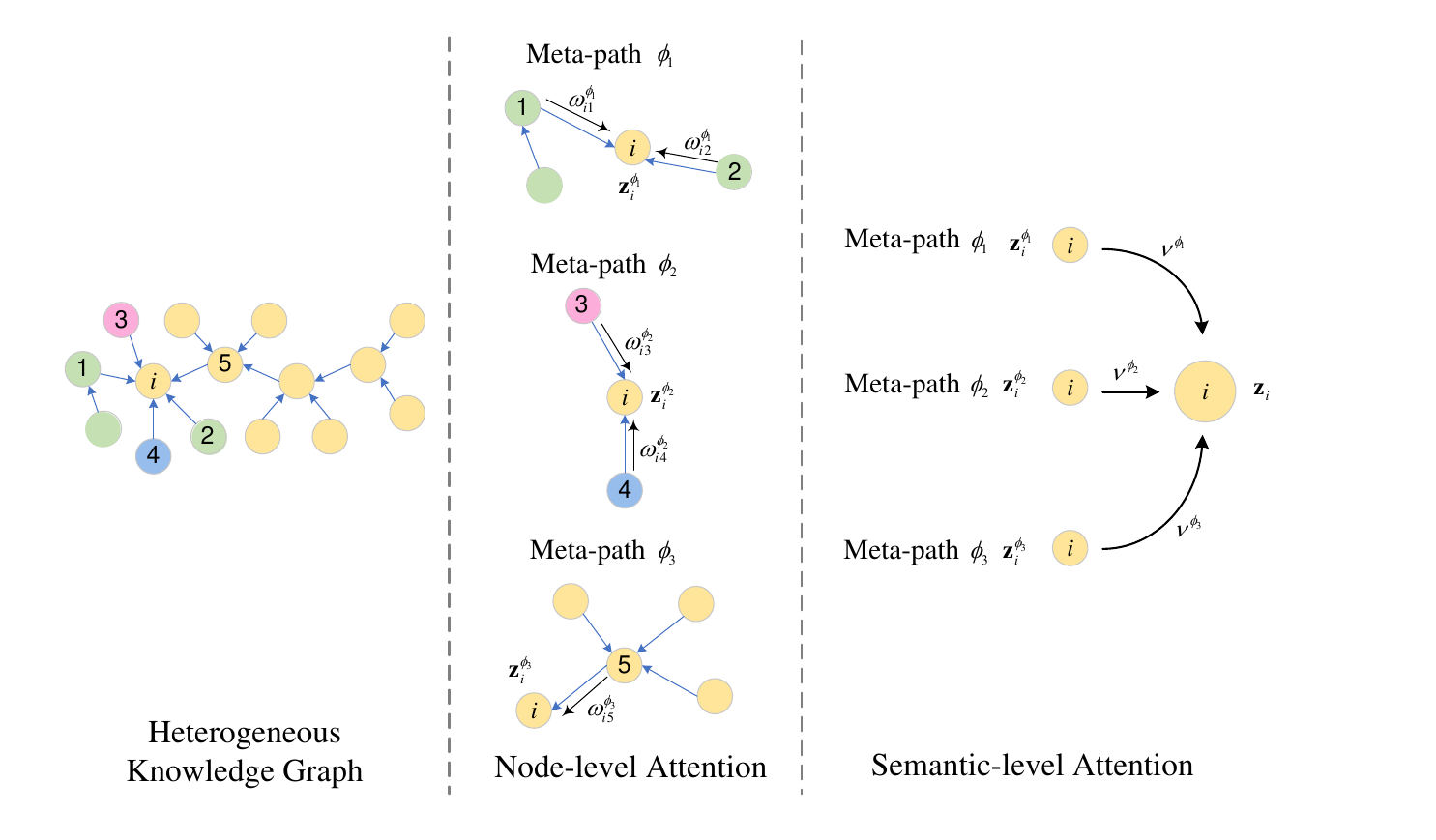}
    \vspace*{-1.5em}
	\caption{Hierarchical attention mechanism.}
	\label{attention_mechanism}
    \vspace*{-1.5em}
\end{figure}

\subsubsection{Temporal convolution layer}
In addition to spatial convolution, temporal convolution is employed to capture the temporal dependencies, thus enabling more comprehensive embeddings. Let $\circledast$ denote the temporal convolution operation and $\Phi\!\in\!\mathbb{R}^{K^\mathcal{S} \!\times\! K^\mathcal{T} \!\times\! c_\mathrm{in}}$ be the $c$-th temporal convolution kernel of the $l$-th layer. The convolution result of the kernel $\Phi$ in the $l$-th layer can be expressed as $\left(\mathcal{H}\right)^{l+1}_{c} \in \mathbb{R}^{(N-K^\mathcal{S}+1) \times (T - K^\mathcal{T} +1)}$. The element of $\left(\mathcal{H}\right)^{l+1}_{c}$ in the $n$-th row and $m$-th column, denoted as $\left(\mathcal{H}\right)^{l+1}_{n,m,c}$, is derived by,
\begin{spacing}{0.7}
\begin{equation}
	\begin{aligned}
		& \left(\mathcal{H}\right)^{l+1}_{n,m,c} = \left(\sigma\left(\Phi  \circledast \mathcal{H}^{l}\right)\right)_{n,m,c}
		\\&=\sigma \left( \sum_{i=0}^{K^\mathcal{S}}\sum_{j=0}^{K^\mathcal{T}}\sum_{k=0}^{c_\mathrm{in}}(\Phi)_{i,j,k}\cdot (\mathcal{H})^{l}_{n+i,m+j,k} \right),
        \quad 1\leq c \leq c_\mathrm{out},
		\label{t-conv}
	\end{aligned}
\end{equation}
\end{spacing}
\noindent where $\sigma$ represents the activation function, and $(\Phi)_{i,j,k}$ and $(\mathcal{H})^{l}_{n+i,m+j,k}$ are the corresponding elements of $\Phi$ and $\mathcal{H}^{l}$, respectively.

According to  \eqref{t-conv}, the temporal convolution results of the $c$-th kernel can be represented as $\left(\mathcal{H}\right)^{l+1}_c$. In the proposed framework, the $l$-th layer contains $c_\mathrm{out}$ convolution kernels. The results of the convolution kernels are concatenated together to form the final output as follows:
\begin{spacing}{0.7}
\begin{equation}
\begin{split}
    \mathcal{H}^{l+1} &= \left[(\mathcal{H})^{l+1}_{1}; (\mathcal{H})^{l+1}_{2}; \ldots; (\mathcal{H})^{l+1}_{c_\mathrm{out}}\right]\\
    &\quad \in \mathbb{R}^{(N-K^\mathcal{S}+1) \times (M-K^\mathcal{T}+1) \times c_\mathrm{out}}.
\end{split}
\end{equation}
\end{spacing}
\noindent Suppose that the total number of layers is $L$, and the representation tensor in the last layer is the final representation matrix, i.e., $\mathcal{H}^L = \mathbf{Z}$.

\subsection{Link Prediction Task}

In general, the quality of a KG embedding algorithm is typically assessed through a link prediction task, where a superior algorithm achieves higher metrics. This subsection details the process of deriving the predicted adjacency matrix $\hat{\mathbf{A}}$ from the final node representation matrix $\mathbf{Z}$. Firstly, node-wise cosine similarity is computed according to
\begin{spacing}{0.7}
\begin{equation}
	{c_{i,j}} = \frac{{\mathbf{z}}_i \cdot {\mathbf{z}_j}}{\Vert \mathbf{z}_i \Vert_2 \cdot \Vert \mathbf{z}_j \Vert_2} ,
\end{equation}
\end{spacing}
\noindent where $c_{i,j}$ represents the cosine similarity between node $i$ and node $j$. Secondly, $c_{ij}$ values are sorted in the descending order and the top-$k$ value is set as the threshold $h$. For each elements in the $\hat{\mathbf{A}}$, $\hat{a}_{i,j}$ is assumed to be 1 when $c_{i,j}$ exceeds the threshold, and set to 0 otherwise, as summarized by (\ref{prediction}):
\begin{spacing}{0.7}
\begin{equation}
	\hat{a}_{i,j} =
	\begin{cases}
		0, & \text{if}\quad c_{i,j} < h\\
		1, & \text{otherwise}.
	\end{cases}
	\label{prediction}
\end{equation}
\end{spacing}
\noindent The representation vector pairs output by each two nodes are subjected to a similarity calculation, and the obtained results are subsequently compared with the graph constructed from expert knowledge. The loss function is designed as follows:
\begin{spacing}{0.7}
\begin{equation}
	\mathcal{L} = \sum_i^N\sum_j^N\left(c_{i,j}-a_{i,j}\right)^2 ,
\end{equation}
\end{spacing}
\noindent where $c_{i,j}$ is the cosine similarity between node pairs and $a_{i,j}$ is the true adjacency matrix elements of wireless data KG. $a_{i,j}$ takes the value of 1 when the two nodes are connected and 0 when they are unconnected. The overall process of STREAM is shown in Algorithm 1.

\begin{algorithm}
\begin{spacing}{1}
	\renewcommand{\algorithmicrequire}{\textbf{Input:}}
	\renewcommand{\algorithmicensure}{\textbf{Output:}}
	\caption{Procedure of STREAM}
	\begin{algorithmic}[1]

    \footnotesize
		\REQUIRE Adjacency matrix $\mathbf{A}$, data matrix $\mathbf{X}$, meta-path set $\{\phi_1, \phi_2, \ldots, \phi_P\}$, maximum training epochs $E$.

		\STATE Initialize the embedding tensor $\mathcal{H}^0 \leftarrow \mathbf{X}$, current epoch $k$ and ST-Conv module number $o$;
		\FOR {$k = \{0, 1, \ldots, K\}$}
		\FOR {$o = \{0, 1\}$}
		
		\STATE Calculate the  $\mathcal{H}^{3o+1}$ by TCN according to Eq. \eqref{t-conv};
		\FOR{$\phi_p \in \{\phi_1, \phi_2, \ldots, \phi_P\}$}
		\STATE Calculate the GCN on  $\mathcal{H}^{3o+1}$ according to Eq. \eqref{gcn};
		\STATE Calculate the node-level coefficient matrix $\mathbf{S}^{1,\phi_p}$ according to Eq. \eqref{omega} and Eq. \eqref{tilde_omega};
		\STATE Obtain $\mathcal{H}^{3o+2, \phi_p}$ by performing the node-level aggregation according to Eq. \eqref{node_level_agg};
		
		\ENDFOR
		\STATE Calculate the meta-path-level coefficient $\{\widetilde{e}^{\phi_1}, \widetilde{e}^{\phi_2}, \ldots, \widetilde{e}^{\phi_P}\}$ according to Eq. \eqref{nu} and Eq. \eqref{tilde_nu};
		\STATE Perform the meta-path-level aggregation according to Eq. \eqref{meta_level_agg}, thus obtaining $\mathcal{H}^{3o+2}$;
		\STATE Calculate the  $\mathcal{H}^{3o+3}$ by TCN according to Eq.  \eqref{t-conv};
		\ENDFOR
		\STATE Embedding matrix $\mathbf{Z}$ is obtained by calculating $\mathcal{H}^6$ through the output layer;
		\STATE Calculate the cosine similarity $c_{i,j}$ and the loss function $\mathcal{L}$;
		\STATE Back propagation and update the network parameters in STREAM;
		\ENDFOR
		\ENSURE  $\mathbf{Z}$.
	\end{algorithmic}
\end{spacing}
	\label{algorithm2}
\end{algorithm}

\subsection{Feature Dataset Generation}

In the above steps, we obtained the cosine similarity between nodes, which can be used to measure the degree of association between them, as shown in (\ref{measureassociation}).
\begin{spacing}{0.7}
\begin{equation}
	\omega_{i,j} =
	\begin{cases}
		0, & \text{if}\quad a_{i,j} = 0 \\
		c_{i,j}, & \text{otherwise}.
	\end{cases}
	\label{measureassociation}
\end{equation}
\end{spacing}

We represent the degree of association between each pair of nodes in the graph  using the matrix $\mathbf{\Omega}$, where $\omega_{i,j}$ is an element of the matrix. At this stage, we can compute the impact of node $v$ on node $u$ in the wireless data KG using (\ref{impactefficiency}), where node $v$ is the $m$-th order neighbor node of node $u$. Here, the $m$-th order neighboring node refers to another node that can be reached by starting from a node and traversing $m$ edges  in the network or graph structure. When $m$ is infinite, it indicates that there is no path connectivity between the two nodes. In the equation, $\prod_{h=1}^{m} \omega_{th}$ represents the product of edge association for all edges on the $t$-th shortest path from node $v$ to node $u$.
\begin{equation}
	i_{vu} =
	\begin{cases}
		\max(\prod_{h=1}^{m} \omega_{th}), & \text{if  $v$ is the $m$-th order neighbor of $u$} \\
		0, & \text{if $v$ is the infinite-order neighbor of $u$}.
	\end{cases}
	\label{impactefficiency}
\end{equation}

Next, we calculate the degree of influence of all nodes on the target KPI, and then sort them. According to the ranking table, we start with the feature ranked highest in importance, using it as the dependent variable to predict the KPI through neural network or similar algorithms. If the predetermined fitting degree is not achieved, the next feature will be added, and the KPI will be predicted again in combination with the first feature. The process stops when the predetermined fitting degree is reached, and continues adding features if the degree is not met, until the goal is achieved. In this way, through the fitness index, we can filter out the most important features as much as possible. These features, namely the relevant nodes in the graph and the data collected by the nodes, are combined to form a feature dataset, which is prepared for input to some intelligent algorithms in the future. The overall process of intelligent generation of the feature dataset is shown in Algorithm 2.

\begin{algorithm}[!t]
    \begin{spacing}{1}
	\renewcommand{\algorithmicrequire}{\textbf{Input:}}
	\renewcommand{\algorithmicensure}{\textbf{Output:}}
	\caption{ Procedure of the intelligent generation of the feature dataset.}

	\begin{algorithmic}[1]
    \small
		\REQUIRE Adjacency matrix $\mathbf{A}$, data matrix $\mathbf{X}$, degree of association matrix $\mathbf{\Omega}$, predetermined fitting degree $d$, target KPI $w$.

		\STATE Initialize an empty importance ranking table $T$.
	    \FOR {each node $u$ in $\mathbf{A}$}
        \FOR {$m = \{0, 1, \ldots, n\}$}

        \STATE Initialize the influence degree $i_{uw}$ of node $u$ on the target KPI $w$ to 0.
		\IF {node $u$ is the $m$-th order neighbor node of target KPI $w$}
        \STATE Compute the influence degree $i_{uw}$ of node $u$ on KPI $w$ according to Eq. (\ref{impactefficiency}).
        \ENDIF
        \STATE Add the influence degree $i_{uw}$ to $T$.
        \ENDFOR
        \ENDFOR
        \STATE Sort $T$ in descending order based on the influence degree.
		
		\STATE Initialize an empty feature dataset $\mathbf{F}$ and an empty  set of selected features $\mathbf{F'}$.
		\FOR {each node $u$ in ranking Table $T$ }
		\STATE Add node $u$ to set $\mathbf{F'}$,
		\STATE Use selected features as the dependent variable to predict the KPI $w$ using neural networks or similar algorithms, obtain the goodness of fit metric $d'$.
		\WHILE {$d'<d$}
        \STATE Select the next node in $T$ for prediction and add it to $\mathbf{F'}$.
		\ENDWHILE
		\ENDFOR
        \STATE Combine the features in $\mathbf{F'}$ with their corresponding data and store them in $\mathbf{F}$.
		\ENSURE $\mathbf{F}$.
	\end{algorithmic}
    \end{spacing}
	\label{algorithm3}
\end{algorithm}

\section{Experimental Results and Analysis}

In this section, we present the specific experimental results of two algorithms: the STREAM and the feature dataset generation algorithm. We begin by comparing STREAM with traditional methods for wireless data KG link prediction tasks. Then, we apply STREAM to a public dataset for traffic flow prediction and compare its performance with classical traffic flow prediction algorithms. Subsequently, we showcase the experimental results of the feature dataset generation algorithm. Lastly, we validate the effectiveness of the feature dataset by comparing it with the original dataset. The main objective of the entire experimental results is to prove that the feature dataset we generated can effectively reduce the training data scale of the network AI model. This is achieved by extracting the minimal yet crucial dataset that mostly impacts the network AI model, ultimately enabling the realization of green and lightweight intelligence.

\subsection{Experiment Settings}

\begin{itemize}
	\item \textbf{Dataset}: To assess the effectiveness of the proposed STREAM, we conduct extensive experiments on wireless data KG with the following settings. We consider a wireless data KG with $M=30$ graph slices, the coherence time $T_\mathrm{c}$ is  100 seconds. To capture the dynamics of a real wireless communication system, data is collected over a 35-minute time interval, yielding a total of 120418-length observations per node. Different from other KGs, the adjacency matrix of wireless data KG is a sparse matrix with 0 and 1 values, where the number of connected edges (denoted by 1) accounting for only 3\% of the total matrix. During the training process, $k$ is set equaling to the number of edges that actually exist in each graph slice.
	\item \textbf{Baseline}: To demonstrate the superiority of STREAM, we compare it with some baselines, including TransE \cite{TransE}, TransH \cite{TransH}, KG2E \cite{KG2E}, and VGAE \cite{VGAE}. Notably, considering that traditional methods ignore the non-negligible information contained in the data matrix $\mathbf{X}$ of wireless data KG, we implemented a pre-training strategy for TransE. Specifically, we initialized the embeddings of TransE with statistical properties of real data, such as minimum, mean, median, etc. In addition, the embedding dimension $c$ is fixed at 128 and consistent across all instances, the remaining bits of its initial embeddings are filled randomly according to an $\mathcal{N}(0, 1)$ distribution. To assess the effectiveness of the hierarchical attention mechanism, we introduced STREAM-homo for comparison. STREAM-homo is a variant of STREAM with the attention mechanism removed. In other words, for STREAM-homo, the graph slices are trained as if they are homogeneous graphs.
	\item \textbf{Training process}: For each graph slice, a fast real-time link prediction is executed. Specifically, the unmasked portion of graph slice is fed into the STREAM, . After a minimal number of epochs (5 in our case) of training, STREAM is capable of predicting the links in the masked portion. In our configurations, the masking proportion is set to $10\%$, and the number of graph slices is 30. The dimension of the convolution kernels are shown in Fig. \ref{model_framework}. Moreover, the batch-size is set to 50 and the number of layer $L$  is 6. The initial learning rate is $10^{-4}$ and it decays by 0.7 every 5 epochs. For the test set, the positive sample consists of the sum of all masked edges (connected edges). To assess the model's performance with extremely unbalanced samples, the number of randomly selected negative test samples (unconnected edges) is set to five times the number of positive samples. This setup allows for a robust evaluation of STREAM's ability to handle imbalanced data.
\end{itemize}

\subsection{Results and Discussions}

\begin{figure}[!t]
    \centering
    \includegraphics[width=3.5in]{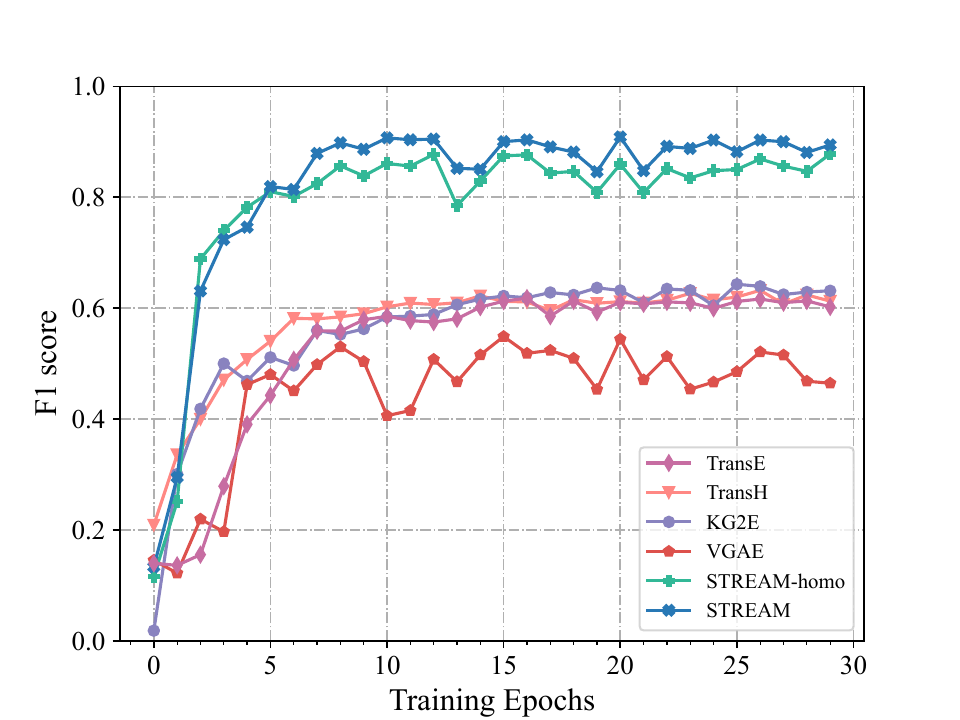}%
    \caption{F1-scores on the training set.}
    \label{training_results}
\end{figure}

Given the uneven distribution of positive and negative samples, relying solely on a single metric like accuracy might not objectively reflect the performance of different algorithms. Therefore, we employ accuracy, precision, recall, F1, and AUC scores to evaluate STREAM. While accuracy, precision, recall and AUC scores are not visualized, F1-scores  for the training set are plotted, and all five metrics can be found in the table for the test set. Fig. \ref{training_results} illustrates the F1-scores on the training set for each graph slice. It is evident that the convergence of both STREAM and STREAM-homo is much faster than that of other baselines. The metric values stabilize after around five epochs, and as the learning rate gradually decreases, fluctuations tend to level off, eventually reaching a relatively stable state. In terms of final convergence values, both STREAM and STREAM-homo outperform other baselines, emphasizing their superiority. Thanks to the hierarchical attention mechanism, STREAM effectively learns the node physical properties in a heterogeneous KG, obtaining more holistic node representation vectors. Consequently, STREAM  marginally outperforms STREAM-homo.


Detailed values are presented in Table \ref{tab:test_results}. In comparison to the baselines, the F1 score of STREAM and STREAM-homo shows an improvement of at least $20\%$. This enhancement is attributed to the ability of our proposed methods to extract information from the synthesis of graph structure, collected data, and heterogeneity. In comparison to STREAM-homo, STREAM still performs approximately  $4\%$ higher compared to STREAM-homo. This difference is due to the intentionally designed hierarchical attention mechanism tailored for heterogeneous graphs in STREAM.

\begin{table}[ht]
	\caption{Simulation Results on Test Set\label{tab:test_results}}
   \footnotesize
	\centering
    \begin{spacing}{1}
	\begin{tabular}{c|ccccc}
		\toprule
    &    Accuracy    &   Precision    &     Recall     &      AUC       &    F1-score\\ \hline
TransE \cite{TransE} &     0.920      &     0.774      &     0.774      &     0.862      &     0.774 \\
TransH \cite{TransH} &     0.933      &     0.811      &     0.811      &     0.885      &     0.811  \\
KG2E  \cite{KG2E} &     0.933      &     0.808      &     0.808      &     0.884      &     0.808   \\
VGAE  \cite{VGAE}  &     0.840      &     0.520      &     0.520      &     0.712      &     0.520  \\
\textbf{STREAM-homo} & \textbf{0.947} & \textbf{0.840} & \textbf{0.840} & \textbf{0.904} & \textbf{0.840}\\
\textbf{STREAM}  & \textbf{0.960} & \textbf{0.880} & \textbf{0.880} & \textbf{0.928} & \textbf{0.880}\\
 \bottomrule

	\end{tabular}
    \end{spacing}
    \vspace*{-1.5em}
\end{table}

\subsection{More Results of Feature Dataset Generation}

\begin{figure}[!t]
    \centering
    \includegraphics[width=3.5in]{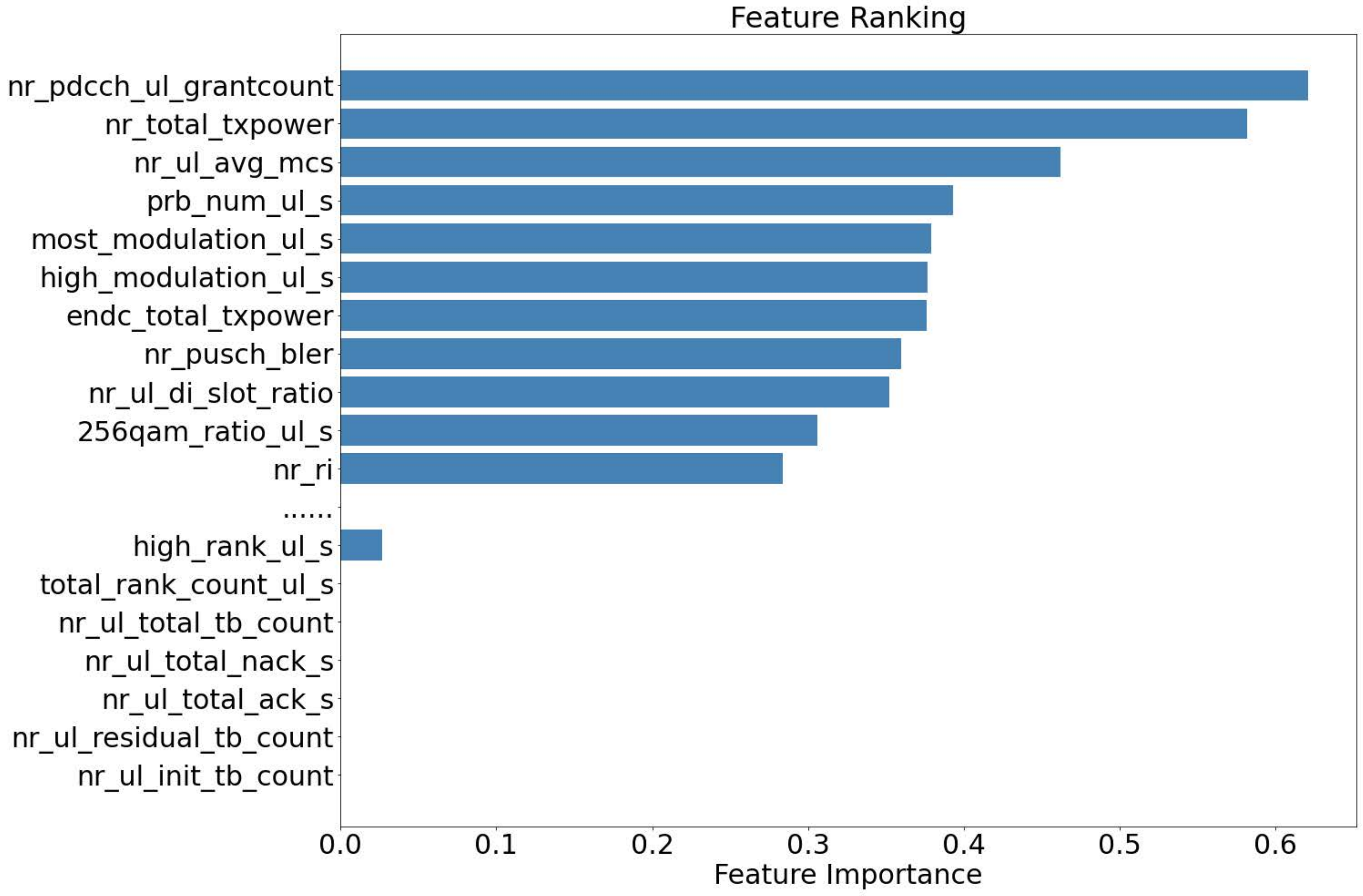}
    \caption{Feature ranking.}
    \label{feature_ranking}
     \vspace*{-1.5em}
\end{figure}

In the initial stage, we curated 82 data fields from a pool of 201, shaping them into a wireless data KG with a focal point on uplink throughput. Subsequently, as depicted in Fig. \ref{feature_ranking}, we executed a sorting process to rank the influence levels of all nodes on the KPI node, specifically targeting the uplink throughput. Due to space limit, we have omitted the middle section of this figure, which includes the influence levels of the remaining nodes on uplink throughput. The prioritization depicted in the figure highlights the significant impact of variables such as user scheduling frequency, power levels, modulation and coding strategies, and the number of uplink physical resource blocks on uplink throughput. These findings, derived from data training, also broadly align with  fundamental principles of communication.

After obtaining the feature ranking table, we set a desired fitting goodness of 0.95 and chose the R2 score as the measurement for fitting goodness. Subsequently, a fully-connected neural network was designed with three hidden layers, each consisting of 32 neurons and utilizing the ReLU activation function. Finally, an output layer was included specifically for predicting the uplink throughput. Following the procedure outlined in Algorithm 2, features were sequentially added to the dependent variable to predict the uplink throughput, until the R2 score surpassed $95\%$. Ultimately, four features were selected: \texttt{nr\_pdcch\_ul\_grantcount}, \texttt{nr\_total\_txpower}, \texttt{nr\_ul\_avg\_mcs}, and \texttt{prb\_num\_ul\_s}. These features yielded an R2 score of $97.36\%$ for predicting uplink throughput. Considering that these features were chosen from a set of 201 data fields, the feature compression rate reached $98.01\%$.  At last, we store the selected features together with the corresponding data for each feature, forming a feature dataset.


\subsection{Benefits of Feature Dataset and its Implications}

The main purpose of this subsection is to evaluate the feature dataset to validate the effectiveness of the proposed PML native AI architecture in achieving green and lightweight intelligence. The quality assessment of the feature dataset primarily depends on its impact on the performance of downstream AI model algorithms. When we can achieve the desired results with minimal key data and computational costs, which previously required a large amount of data and computational expenses, it demonstrates that this architecture is a viable approach for achieving green and lightweight intelligence.

Upon obtaining the feature dataset from the wireless data KG, it brings several advantages. Firstly, regarding the KPI of uplink throughput, the original dataset comprising 201 data fields has been streamlined to only 4 data fields. This drastic reduction eliminates extraneous nodes, enabling subsequent research on uplink throughput in real network environments to concentrate on essential data fields. Secondly, intelligent communication systems incur additional bandwidth allocation for data transmission. Due to limited bandwidth, the quantity of data to be transmitted is restricted. Therefore, it is necessary to employ a feature dataset that conveys maximum information while minimizing its size, thus facilitating efficient data transmission. Lastly, real-time intelligence in wireless networks necessitates minimizing computational costs to avoid latency and energy wastage.
In order to predict the throughput fairly, we removed all data fields in the throughput class when inputting features and used the remaining 188 features to predict the physical layer uplink throughput. Based on the experimental results in Table \ref{cost}, we were able to achieve an excellent fit of $99.97\%$. The results of training the model on the feature dataset show similar performance in comparison, but the number of features is reduced by about $97.9\%$, the number of parameters is reduced by about $71.87\%$, and the floating point operations (FLOPs) and execution time are both reduced by almost an order of magnitude. These results indicate a significant reduction in computational overhead, providing preliminary support for the subsequent implementation of green intelligence.

\begin{table}
\begin{spacing}{1.2}
\begin{center}
\caption{Performance and Cost Comparison of AI Models based on Raw Dataset and Feature Dataset}
\label{cost}
\footnotesize
\begin{tabular}{| c | c | c |}

\hline
 & AI Models based & AI Models based \\
 & on Raw Dataset & on Feature Dataset \\
\hline
Number of feature    & 188                             & 4 \\
\hline
Fitting dgree        &  99.97\%                        & 97.36\% \\
\hline
Model parameters     & 8193                            & 2305 \\
\hline
FLOPs (G)            &     $1.63 \times 10^{-5} $      &   $ 4.51\times 10^{-6} $         \\
\hline
Execution time (s)   &     465.75                      &    28.33         \\
\hline
\end{tabular}
\end{center}
\end{spacing}
\vspace*{-1.5em}
\end{table}


\section{Conclusion}

In this paper, we proposes a PML native AI architecture for green intelligent communications. This architecture incorporates KGs into the field of wireless communication, forming a wireless data KG, and utilizes it to generate feature datasets on demand. This provides a feasible path for achieving green, lightweight real-time intelligent communications. To improve the efficiency of wireless data KG construction, the STREAM is proposed. STREAM aims to improve the utilization of real-world wireless big data and expert knowledge, automating the completion and intelligent construction of the wireless data KG. Compared to other algorithms, STREAM exhibits outstanding performance in F1 and AUC scores when predicting hidden relationships. Furthermore, after obtaining the degree of correlation between nodes through the STREAM, it is possible to further explore the relationships and graph structure among these nodes, enabling the deep mining of the minimal and most effective feature dataset that influences the target KPI. This feature dataset reduces the training overhead of the AI model by almost an order of magnitude and provides a valuable reference for the input of the AI model. Future research will continue to follow this architecture, using the generated feature dataset to drive the training of AI models in specific application scenarios, promoting further advancements in this field.




\begin{thebibliography}{10}
\providecommand{\url}[1]{#1}
\csname url@samestyle\endcsname
\providecommand{\newblock}{\relax}
\providecommand{\bibinfo}[2]{#2}
\providecommand{\BIBentrySTDinterwordspacing}{\spaceskip=0pt\relax}
\providecommand{\BIBentryALTinterwordstretchfactor}{4}
\providecommand{\BIBentryALTinterwordspacing}{\spaceskip=\fontdimen2\font plus
\BIBentryALTinterwordstretchfactor\fontdimen3\font minus
  \fontdimen4\font\relax}
\providecommand{\BIBforeignlanguage}[2]{{%
\expandafter\ifx\csname l@#1\endcsname\relax
\typeout{** WARNING: IEEEtran.bst: No hyphenation pattern has been}%
\typeout{** loaded for the language `#1'. Using the pattern for}%
\typeout{** the default language instead.}%
\else
\language=\csname l@#1\endcsname
\fi
#2}}
\providecommand{\BIBdecl}{\relax}
\BIBdecl

\bibitem{ITU2023}
International Telecommunication Union, ``IMT Vision-Framework and overall objectives of the future development of IMT for 2030 and beyond,'' Recommendation ITU-R M.2160-0, Nov. 2023.

\bibitem{You2021Wang}
X.~You, C.-X.~Wang, et al., ``Towards 6G wireless communication networks: Vision, enabling technologies, and new paradigm shifts,''  \emph{Sci. China Inf. Sci.}, vol.~64, no.~1, pp. 1--74, Jan. 2021.


\bibitem{Masood2023Farooq}
U.~Masood, H.~Farooq, A.~Imran, and A.~Abu-Dayya, ``Interpretable AI-Based large-scale 3D pathloss prediction model for enabling emerging self-driving networks,''  \emph{IEEE Trans. Mob. Comput.}, vol.~22, no.~7, pp. 3967--3984,  Jul. 2023.

\bibitem{Letaief2019Chen}
K.~B.~Letaief, W.~Chen, Y.~Shi, et al., ``The roadmap to 6G: AI empowered wireless networks,''  \emph{IEEE Commun. Mag.}, vol.~57, no.~8, pp. 84--90, 2019.

\bibitem{Chen2020Liu}
Y.~Chen, W.~Liu, Z.~Niu, et al., ``Pervasive intelligent endogenous 6G wireless systems: Prospects, theories and key technologies,'' \emph{Digital Communications and Networks}, vol.~6, no.~3, pp. 312--320, 2020.

\bibitem{IEA}
International Energy Agency, ``Net Zero by 2050,'' [Online]. Available: https://www.iea.org/reports/net-zero-by-2050, Jun. 2021.

\bibitem{Huang2019Yang}
T.~Huang, W.~Yang, J.~Wu, et al., ``A survey on green 6g network: architecture and technologies,'' \emph{IEEE Access}, vol. 7, pp. 175758--175768, Dec. 2019.

\bibitem{Polese2021Jana}
M.~Polese, R.~Jana, V.~Kounev, K.~Zhang, S.~Deb, and M.~Zorzi, ``Machine learning at the edge: A data-driven architecture with applications to 5G cellular networks,'' \emph{IEEE Trans. Mob. Comput.}, vol.~20, no.~12, pp. 3367--3382, Dec. 2021.

\bibitem{You2023Huang}
X.~You, Y.~Huang, et al., ``Toward 6G $\text{TK}\upmu$ extreme connectivity: Architecture, key technologies and experiments,'' \emph{IEEE Wirel. Commun.}, vol. 30, no. 3, pp. 86--95, Jun. 2023.


\bibitem{Xu2023}
W.~Xu, et al., ``Edge learning for B5G networks with distributed signal processing: Semantic communication, edge computing, and wireless sensing,'' \emph{IEEE J. Sel. Topics Signal Process.}, vol. 17, no. 1, pp. 9--39, Jan. 2023.

\bibitem{Xu2018Xu}
W.~Xu, Y.~Xu, C.~-H. Lee, Z.~Feng, P.~Zhang, and J.~Lin, ``Data-cognition-empowered intelligent wireless networks: Data, utilities, cognition brain, and architecture,'' \emph{IEEE Wirel. Commun.}, vol.~25, no.~1, pp. 56--63, Feb. 2018.

\bibitem{Liu2024AAAI}
S.~Liu, X.~Li, Z.~Mao, P.~Liu, and Y.~Huang, ``Model-driven deep neural network for enhanced AoA estimation using 5G gNB,'' in \emph{Proc. 38th Annu. AAAI Conf. Artificial Intell. (AAAI)}, Vancouver, BC, Canada, 2024, pp. 10775.

\bibitem{Liu2020Bi}
Y.~Liu, S.~Bi, Z.~Shi, and L.~Hanzo, ``When machine learning meets big data: A wireless communication perspective,'' \emph{IEEE Veh. Technol. Mag.}, vol.~15, no.~1, pp. 63--72, Mar. 2020.


\bibitem{Ding2022Lai}
S.~Ding, Q.~Lai, Z.~Zhou, J.~Gong, J.~Cui, and S.~Liu, ``A novel deep learning model for link prediction of knowledge graph,'' in \emph{Proc. IEEE Int. Symp. Circuits Syst. (ISCAS)}, Austin, TX, USA, 2022, pp. 2477--2481.

\bibitem{Shen2022Zhang}
Y.~Shen, J.~Zhang, S. H.~Song, and K. B.~Letaief, ``Graph neural networks for wireless communications: From theory to practice,'' \emph{IEEE Trans. Wirel. Commun.}, vol. 22, no. 5, pp. 3554--3569, Nov. 2022.

\bibitem{Wang2017Mao}
Q.~Wang, Z.~Mao, B.~Wang, and L.~Guo, ``Knowledge graph embedding: A survey of approaches and applications,'' \emph{IEEE Trans. Knowl. Data En.}, vol. 29, no. 12, pp. 2724-2743, Dec. 2017.

\bibitem{TTIN}
Y.~Huang, S.~Liu, C.~Zhang, X.~You, and H.~Wu, ``True-data testbed for 5G/B5G intelligent network,'' \emph{Intell. Converged Networks}, vol. ~2, no.~2, pp. 133--149, Jun. 2021.

\bibitem{3GPP}
3GPP. ``Summary of Rel17 Work Items,'' 3GPP TR 21.205, V1.1.0, 2023.  [Online]. Available: https://www.3gpp.org/ftp/Specs/archive/21\_series/21.205.

\bibitem{Wang2019Ji}
X.~Wang, H.~Ji, C.~Shi, et al, ``Heterogeneous graph attention network,'' in \emph{Proc. World Wide Web Conf. (WWW)}, San Francisco, CA, United states, 2019, pp. 2022--2032.

\bibitem{TransE}
A.~Bordes, N.~Usunier, A.~Garcia-Duran, et al, ``Translating embedding for modeling multi-relational data,'' in \emph{Proc. 26th Annu. Conf. Neural Inf. Proces. Syst. (NIPS)}, Lake Tahoe, NV, USA, 2013, pp. 2787--2795.

\bibitem{TransH}
Z.~Wang, J.~Zhang, J.~Feng, et al, ``Knowledge graph embedding by translating on hyperplanes,'' in \emph{Proc. 28th AAAI Conf. Artif. Intell. (AAAI)}, Qu\'{e}bec City, QC, Canada, 2014, pp. 1112--1119.

\bibitem{KG2E}
S.~He, K.~Liu, G.~Li, et al, ``Learning to represent knowledge graphs with Gaussian embedding,'' in \emph{Proc. 24th ACM Int. Conf. Inf. Knowledge Manage. (CIKM)}, Melbourne, VIC, Australia, 2015, pp. 623--632.

\bibitem{VGAE}
T. N. Kipf, M. Welling, ``Variational graph auto-encoders,'' in \emph{arXiv preprint arXiv:1611.07308}, 2016.




\end{thebibliography}
\end{document}